\documentclass[12pt]{article}
\usepackage{epsf}

\usepackage{amssymb}
\usepackage{amsmath}
\usepackage{graphicx}
\usepackage{latexsym}

\begin{document}

\def\ds{\displaystyle}
\def\Slash#1{{\ooalign{\hfil$#1$\hfil\crcr\hfil$/$\hfil}}}

\newcommand{\rem}[1]{{\bf #1}}

\renewcommand{\theequation}{\thesection.\arabic{equation}}

\renewcommand{\thefootnote}{\fnsymbol{footnote}}
\begin{titlepage}

\def\thefootnote{\fnsymbol{footnote}}

\begin{center}

%
\vskip .75in

{\Large \bf 
  New physics for muon anomalous magnetic moment and its electroweak
     precision analysis
}
\vskip .75in

{\large
$^{(a)}$Shunichi Kanemitsu and $^{(a, b)}$Kazuhiro Tobe
}

\vskip 0.25in

{\em $^{(a)}$Department of Physics, Nagoya University,
Nagoya 464-8602, Japan}

{\em $^{(b)}$Kobayashi-Maskawa Institute for the Origin of Particles and
 the Universe, \\Nagoya University,
Nagoya 464-8602, Japan}

\end{center}
\vskip .5in

\begin{abstract}

 About 3$\sigma$ deviation from the standard model prediction of muon
 anomalous magnetic moment (muon g-2) has been reported. We consider new
 physics beyond the standard model which has new Yukawa interactions with
 muon. We compute new contributions to muon g-2 and corrections to
 electroweak observables, and show the consistent region of parameter
 space. We find that in a simple model where the chirality flip of muon
occurs  only in the external muon line in one-loop muon g-2 diagrams, 
it is necessary to introduce the relatively
 large new Yukawa coupling and the electroweak scale new particles. 
 On the other hand, in a model where
 the chirality flip can occur in the internal fermion line of one loop 
muon g-2 diagrams, we can obtain
 favorable g-2 contributions without large Yukawa coupling, and
 they are consistent with the precision electroweak observables.
 Finally, we discuss effects of new particles for muon g-2 on the Higgs boson decay
$h\rightarrow \gamma \gamma$ and direct productions of these particles at the LHC
 experiment. 

 \end{abstract}

\end{titlepage}

\renewcommand{\thepage}{\arabic{page}}
\setcounter{page}{1}
\renewcommand{\thefootnote}{\#\arabic{footnote}}

\section{Introduction}
\label{sec:intro}
\setcounter{equation}{0}

The standard model of elementary particles (SM) has been amazingly successful,
and currently the Large Hadron Collider (LHC) experiment is searching
for Higgs boson, which is the only particle that has not been
observed yet in the framework of the SM. So far, some interesting hints
of the Higgs boson with a mass of about 125 GeV have been reported at
the LHC~\cite{ATLAS:2012ae} and 
its discovery may not be in a far future.
The Higgs boson with a mass of about 100 GeV is totally consistent
with the electroweak (EW) precision measurements.
Therefore, the SM well-describes the nature up
to the EW scale. 

We, however, think that the SM is not the final theory of elementary
particles. The hierarchy problem has been a strong driving force to
think of the physics beyond the SM, and many ideas, such as
supersymmetry, extra-dimension, and technicolor,  have been proposed to
solve the problem. The LHC experiment is an ideal place to probe the
models for the hierarchy problem, and the searches are currently going
on. The first round of the LHC searches, however, does not show any
serious deviations from the SM prediction. It seems that this
negative search result starts creating a tension between new
physics models and the requirements for the hierarchy problem. Therefore
it will be good time to consider other approaches to think about
new physics beyond the SM.

There is a different approach to consider the physics beyond
the SM, based on a consideration for unexplained experimental results.  
The muon anomalous magnetic moment (muon g-2) is one of the most
precisely measured
observables~\cite{Nakamura:2010zzi}. 
The theoretical prediction from the SM
by several groups has suggested that there is a discrepancy between the 
experimental result and the SM prediction~\cite{Hagiwara:2011af}: 

\begin{eqnarray}
\delta a_\mu \equiv a_\mu^{\rm exp} -a_\mu^{\rm SM}=(26.1\pm 8.0) \times
 10^{-10},
\end{eqnarray}
where $\delta a_\mu$ is a discrepancy between the experimental result
$(a_\mu^{\rm exp})$ and the SM prediction $(a_\mu^{\rm SM})$.
There are many
discussions~\cite{Hagiwara:2011af,Boughezal:2011vw}
on uncertainties of the SM prediction to
understand whether this anomaly is real or not. 
If the discrepancy can not be explained by the SM, this
would be an evidence of the physics beyond the SM. At present, there
seems no satisfactory explanation for this anomaly within the SM. 
Therefore, it is worth while considering the physics beyond the SM
seriously in order to explain the anomaly.

It is interesting to note that the anomaly of muon g-2 is the same size as
1-loop contributions induced by the EW gauge bosons in the SM.
This suggests that in order to explain the anomaly, new particles
with masses of EW scale are required if the interactions are of order
of EW gauge couplings. 

Although the anomaly of muon g-2 has been discussed in the context of new physics models
(for example, see Ref.~\cite{Moroi:1995yh} in MSSM, Ref.~\cite{McLaughlin:2000zf} in extra dimension model, and 
Ref.~\cite{Choudhury:2006sq} in Little Higgs model) motivated mainly by the hierarchy
problem, we take a ``bottom-up'' style approach \footnote{ See also
Ref.~\cite{Kannike:2011ng} for the recent related studies.}.
Starting from introducing new interactions 
with muon to generate new contribution to muon g-2, we try to capture important features
of new physics models, although we can only discuss a part of complete models.
In this paper, we consider models where muon has new Yukawa type interactions with 
new particles to generate new contributions to muon g-2.
We find that in a model where the chirarity flip of muon occurs only in
external line of muon g-2 diagrams, it is necessary to
introduce the relatively large Yukawa coupling and the EW
scale new particles. On the other hand, in a model where the chirality
flip can happen in the internal lines of muon g-2 diagrams, we can
obtain favorable g-2 contributions without large Yukawa couplings, and
they are also consistent with the precision EW observables.
We also expect these
particles may be observed directly and/or indirectly at the LHC experiment.

This paper is organized as follows. In the next section, we consider
new physics models where muon has new Yukawa couplings to explain
the anomaly of muon g-2. One model is that only right-handed muon
has new Yukawa coupling. Another one is that both right- and left- 
handed muons have new Yukawa interactions to enhance the contributions
to the muon g-2. We show the consistent region of parameter space.
In Sec.~3, we discuss effects of the new physics on EW
observables and whether the parameter space which is consistent with 
the muon g-2 is also consistent with the precision measurements.
In Sec.~4, we discuss phenomenology of these new physics models
at the LHC. Especially we show the possible effect on the Higgs boson
decay of $h\rightarrow \gamma \gamma$, and the direct production cross
sections of new particles at the LHC. Sec.~5 is devoted to summary.

\section{New physics models for anomaly of muon g-2}

In order to explain the deviation from the SM prediction 
of muon g-2 by new physics, muon has to have new interactions
with some charged particles. Since the effective operator for muon g-2
couples to photon, the loop effects via the new interactions of muon 
with charged particles could induce the extra contribution to the muon g-2.

In this paper, as such new interactions, we consider 
new Yukawa-type couplings with right-handed or/and left-handed muon.
We consider two cases: (1) Model where only right-handed 
muon has new Yukawa interaction and (2) Model where both
right-handed and left-handed muons have new Yukawa interactions.

\subsection{Model where right-handed muon has new
 Yukawa interaction with SU(2) singlet scalar ($\phi$)
 and singlet Dirac fermion ($\chi$)}

We consider the following Yukawa interaction where the right-handed muon
couples to new SU$(2)_L$ singlet fermion $\chi$ and singlet scalar $\phi$:
\begin{eqnarray}
{\cal L}=-y_N \bar{\mu}_R \chi_L \phi-m_\chi\bar{\chi}_R\chi_L+{\rm
 h.c.}-m_\phi^2 \phi^\dagger\phi+{\cdots}.
\end{eqnarray}
Here $\mu_R$ is the right-handed muon. QED charges of new fermion $\chi$ 
and scalar $\phi$ are $Q_\chi$ and $Q_\phi=-1-Q_\chi$, respectively, in
order to have the gauge invariant Yukawa interaction.
The masses of $\chi$ and $\phi$ are denoted by $m_\chi$ and $m_\phi$, respectively.
In order to simplify the model, we have assumed a $Z_2$ parity, under which
the SM particles are even and new particles $\phi$ and $\chi$ are odd. This type of
the parity may be interesting for the dark matter. We also note that 
even if we impose the $Z_2$ parity, the right-handed electron, for example, can have similar
Yukawa interaction with $\phi$ and $\chi$. This will cause severe flavor mixing problem.
We will not discuss the complete model here, however, we implicitly assume that 
$\chi$ or $\phi$ has approximate muon flavor number so that the flavor mixing is
strongly suppressed.

\begin{figure}[ht]
\centerline{
\includegraphics[width=15.0cm,angle=0]{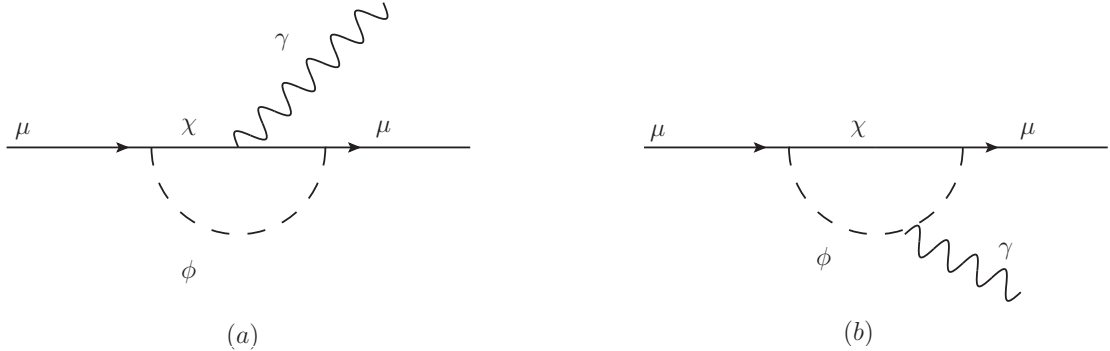}
}
\caption{Feynman diagrams for muon $g-2$}
\label{diagramg2}
\end{figure}
Since the right-handed muon couples to $\chi$ and $\phi$, their
radiative corrections induce the muon g-2, 
$a_\mu^{\rm new}$ as shown in Fig.~\ref{diagramg2}. The contribution $a_\mu^{\rm new}$ 
is given by
\begin{eqnarray}
a_\mu^{\rm new} &=& -\frac{y_N^2m_\mu^2}{16\pi^2}\left[
Q_\chi(C_{11}+C_{21})(\phi,\chi,\chi;p,-q)
-Q_\phi
(C_{12}+C_{22})(\phi,\phi,\chi;q,p-q)
\right],
\label{muong2}
\end{eqnarray}
where $p$ and $p-q$ are momenta of external muons and $q$ is a photon momentum and
a limit $q^2\rightarrow 0$ is taken.
Here $C_X(A,B,C;p_1,p_2)~(X=11,~21,~12,~22)$ are so called Passarino-Veltman functions
\cite{'tHooft:1978xw}, and their definitions in this paper are shown in Appendix. The
explicit forms are
\begin{eqnarray}
(C_{11}+C_{21})(\phi,\chi,\chi;p,-q)&=&\frac{1}{m_\phi^2}\frac{
2-3y-6y^2+y^3+6y\log y}{6(1-y)^4},\\
(C_{12}+C_{22})(\phi,\phi,\chi;q,p-q) &=&\frac{1}{m_\phi^2}
\frac{1-6y+3y^2+2y^3-6y^2\log y}{6(1-y)^4},
\end{eqnarray}
where $y=m^2_\chi/m^2_\phi$, $q^2=0$ and the higher order terms of
$O(m_\mu^2/m_\phi^2)$ are neglected.
The first term in Eq.~(\ref{muong2})
comes from Fig.~\ref{diagramg2}(a) and the second is from Fig.~\ref{diagramg2}(b).

\begin{figure}[ht]
\centerline{
\includegraphics[width=12.0cm,angle=0]{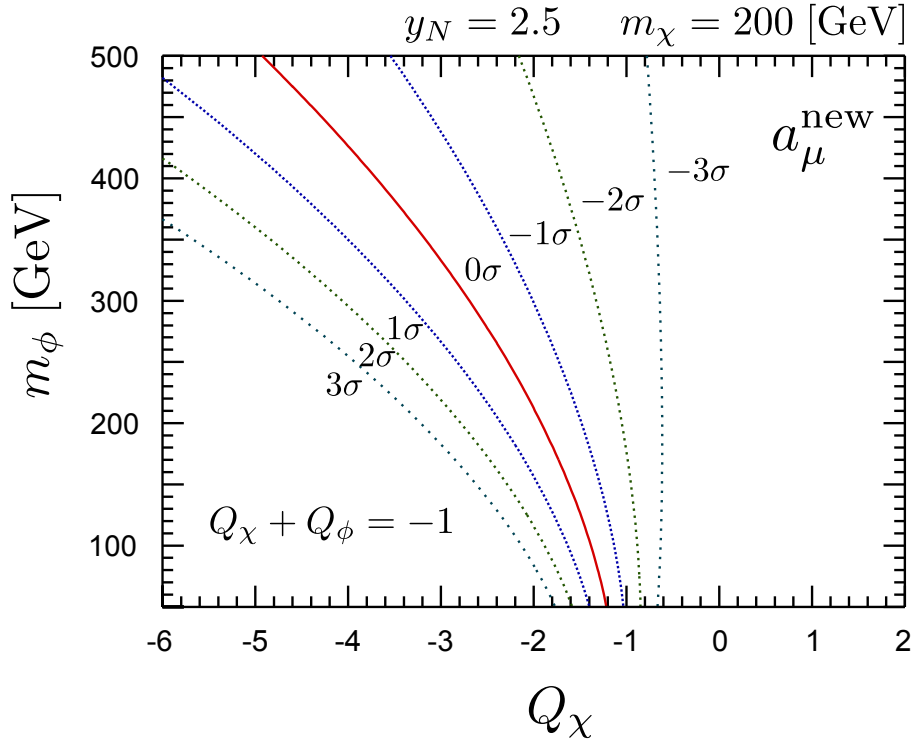}
}
\caption{New physics contribution to muon g-2 ($a_\mu^{\rm new}$)
as a function of $Q_\chi$ and $m_\phi$. Here we take $y_N=2.5$ and $m_\chi=200$ GeV.
We show contours 
for
$(a_\mu^{\rm new}/10^{-10})=2.1,~10.1,~18.1,~26.1,~34.1,~42.1$ and 
$50.1$, from right to left, corresponding to 
$-3\sigma,~-2\sigma,~-1\sigma,~0\sigma,~1\sigma,~
2\sigma$ and $ 3\sigma$ deviation from the measured value, respectively.}
\label{fig_amu}
\end{figure}

In Fig.~\ref{fig_amu}, we show the contribution to $a_\mu^{\rm new}$
as a function of $Q_\chi$ and $m_\phi$.
Here we take $y_N=2.5$ and $m_\chi=200$ GeV. 
We show contours for
$(a_\mu^{\rm new}/10^{-10})=2.1,~10.1,~18.1,~26.1,~34.1,~42.1$ and 
$50.1$, from right to left, corresponding to 
$-3\sigma,~-2\sigma,~-1\sigma,~0\sigma,~1\sigma,~
2\sigma$ and $ 3\sigma$ deviation from the measured value, respectively.
We note that $y_N$ dependence of $a_\mu^{\rm new}$ is trivial, that is, 
$a_\mu^{\rm new}$ is proportional to $y_N^2$, as shown in Eq.~({\ref{muong2}}).
For example, if one takes $y_N=1$, the values of $a_\mu^{\rm new}$ in Fig.~\ref{fig_amu} 
reduce by a factor $(1/2.5)^2=0.16$.

One can see that the region with $Q_\chi>-1$ (which corresponds to $Q_\phi<0$)
is disfavored by the data of muon g-2.
It is interesting to notice that the neutral
fermion ($Q_\chi=0$, corresponding to $Q_\phi=-1$) is difficult to explain the
anomaly in any values of $m_\phi$, and on the other hand, the neutral scalar 
($Q_\phi=0$, corresponding to $Q_\chi=-1$) can
potentially accommodate the anomaly if the scalar $\phi$ is not so
heavy. It is also interesting to note that the multi-charged fermion and scalar
(such as $Q_\chi=-2,~-3,\cdots$, corresponding to $Q_\phi=1,~2,\cdots$)
are also favored by the muon g-2 anomaly if $m_\phi$ is a right value for 
the anomaly. Therefore, the anomaly of muon g-2  constrains the QED charges as well 
as the mass scale of new particles.

\begin{figure}[ht]
\centerline{
\includegraphics[width=8.5cm,angle=0]{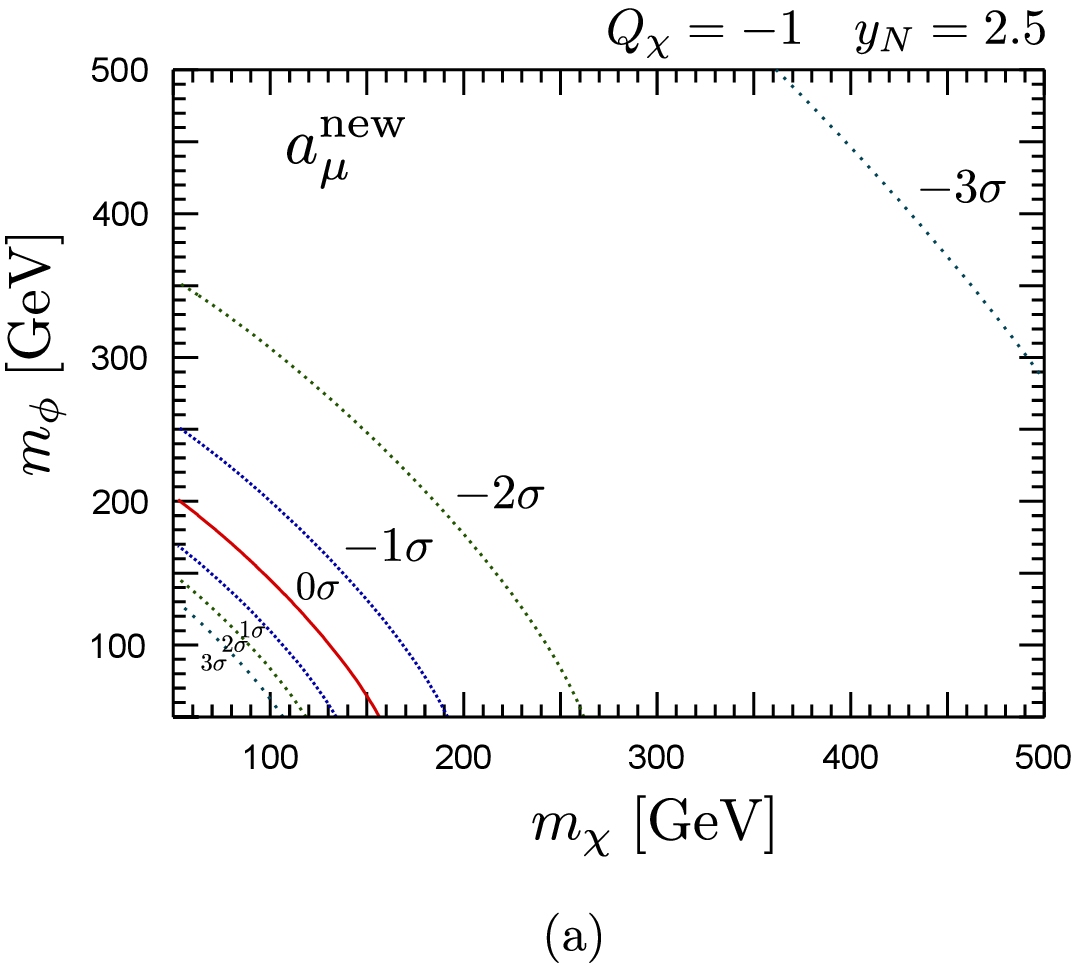}
\includegraphics[width=8.5cm,angle=0]{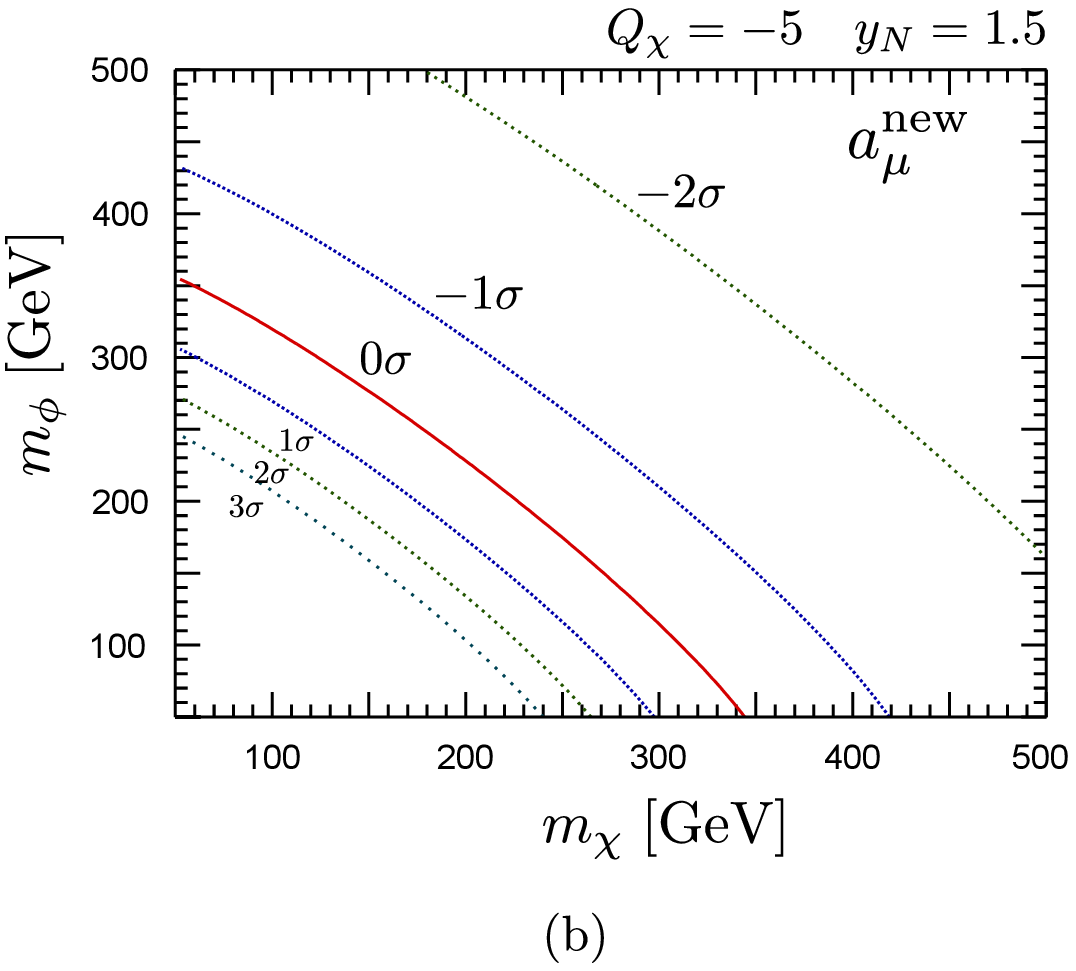}
}
\caption{New physics contribution to muon g$-$2 ($a_\mu^{\rm new}$) 
as a function of $m_\chi$ and $m_\phi$ for (a) $Q_\chi=-1$ and and $y_N=2.5$
(b) $Q_\chi=-5$ and $y_N=1.5$, respectively.
Contours for $(a_\mu^{\rm new}/10^{-10})=2.1,~10.1,~18.1,~26.1,~34.1,~42.1$ and 
$50.1$ (corresponding to 
$-3\sigma,~-2\sigma,~-1\sigma,~0\sigma,~1\sigma,~
2\sigma$ and $ 3\sigma$ deviation from the measured value) are shown, similar to
Fig.~\ref{fig_amu}.
}
\label{fig_amu_2}
\end{figure}

In Fig.~\ref{fig_amu_2}, we show $a_\mu^{\rm new}$ as a function of
$m_\chi$ and $m_\phi$ for (a) $Q_\chi=-1$ $({\rm corresponding~to}~Q_\phi=0)$ 
and $y_N=2.5$ and (b) $Q_\chi=-5$ $({\rm corresponding~to}~Q_\phi=4)$ and $y_N=1.5$.
Contours for $(a_\mu^{\rm new}/10^{-10})=2.1,~10.1,~18.1,~26.1,~34.1,~42.1$ and 
$50.1$ (corresponding to 
$-3\sigma,~-2\sigma,~-1\sigma,~0\sigma,~1\sigma,~
2\sigma$ and $ 3\sigma$ deviation from the measured value) are shown, similar to
Fig.~\ref{fig_amu}.\footnote{
If one changes the value of Yukawa coupling $y_N$, one should simply multiply
the value of $a_\mu^{\rm new}$ in Fig.~\ref{fig_amu_2} (a) and (b) by
a factor $(\frac{y_N}{2.5})^2$ and $(\frac{y_N}{1.5})^2$, respectively.}
As one can see, new particles with their masses of about $100$ GeV and
the coupling $y_N\sim O(1)$ are strongly favored by the anomaly of muon g-2
when $Q_\chi=-1$. When the QED charge of $\chi$ decreases further negatively
(from $-1$ to $-5$), 
the muon g-2 gets larger even if the Yukawa coupling $y_N$ gets smaller, 
as shown in Fig.~\ref{fig_amu_2} (b).

Since the anomaly of muon g-2 requires relatively light new particles
($\sim O(100~{\rm GeV})$) and/or relatively large Yukawa coupling
($\sim O(1)$) in this scenario, the EW precision
observables can be affected by these new particles and new interactions. 
Therefore, we will
check the effects of these particles on the EW observables in a next section.

We would also like to point out that the anomaly of muon g-2 suggests
the existence of relatively light new particles and/or multi-charged particles. 
Therefore, it is very interesting to know whether these particles can be found 
directly or indirectly at the LHC. We will study the direct productions of these 
particles and the effects on Higgs decay to $\gamma \gamma$ at the LHC in a 
later section. 

\subsection{ Model where both right- and left-handed muons have new
  Yukawa couplings 
}
Unlike the previous case, here we consider that 
both right- and left-handed muons have new Yukawa interactions:
\begin{eqnarray}
{\cal L} &=& -y_L \bar{L}_2 \Phi \chi_R -y_R \bar{\mu}_R \phi \chi_L
 -m_\chi \bar{\chi}_L \chi_R+
{\rm h.c.},
\end{eqnarray}
where $L_2~(=(\nu_{\mu L},~\mu_L)^{\rm T})$, $\Phi~(=(\phi_1,\phi_2)^{\rm T})$ and $\phi$ are 
the second generation SU(2) doublet lepton, SU(2) doublet and singlet scalars,
respectively, and $\chi$ is SU(2) singlet fermion, whose mass is  $m_\chi$.
Here we have assumed the $Z_2$ parity again in order to simplify the model. Under the $Z_2$
parity, the SM particles are even, and new particles $\phi$, $\Phi$ and $\chi$ are odd.
We also implicitly assume the approximate muon flavor symmetry, under which $\chi$
or ($\phi,~\Phi$) has muon flavor number, so that the flavor mixing of this type
of new Yukawa couplings are strongly suppressed. The model we discuss in this paper
may be a part of the complete model. However, we think this part of the complete model
is crucial for the muon g-2.\footnote{In summary, we briefly comment on the possible
completions of this type of model.}

The QED charges of new particles are represented in term of QED charge of $\chi$
field ($Q_\chi$) as follows: 
\begin{eqnarray}
Q(\phi_1)&\equiv& Q_1=-Q_\chi,\\
Q(\phi_2)&\equiv& Q_2 = -1-Q_\chi,\\
Q_\phi&=&-1-Q_\chi=Q_2.
\end{eqnarray}
Since $Q_\phi=Q_2$, $\phi$ and $\phi_2$ can mix each other. 
For example, the following gauge invariant term
induces the $\phi_2-\phi$ mixing mass terms after the electroweak symmetry
is spontaneously broken:
\begin{eqnarray}
{\cal L} = -\lambda M (H^\dagger \Phi \phi^\dagger)+{\rm h.c.}=-\frac{\lambda M v}{\sqrt{2}}
\phi_2\phi^\dagger+\cdots,
\end{eqnarray}
Here we parameterize the mass terms for $\phi$
and $\phi_2$ as follows:
\footnote{When $Q_\phi=0$, mass term such as $-m^2\phi\phi_2$ is also possible.
Here we neglect such a mass term for simplicity. Even if we include such a mass term,
our result does not change qualitatively.
}
\begin{eqnarray}
{\cal L}=-\left(\phi^{\dagger},~\phi_2^{\dagger}
\right) \left(
\begin{array}{cc}
m^2_{11} &  m^2_{12}\\
m^2_{12} & m^2_{22}
\end{array}
\right)\left(
\begin{array}{c}
\phi\\
\phi_2
\end{array}
\right).
\label{mass_matrix}
\end{eqnarray}
Thus diagonalizing this mass matrix, we define the mass eigenstates $s_i~(i=1,2)$ as
\begin{eqnarray}
\left(
\begin{array}{c}
\phi
\\
\phi_2
\end{array}
\right)_i= V_{ij}s_j.
\end{eqnarray}
Here $V_{ij}$ is a mixing unitary matrix, which diagonalizes the mass
matrix shown in Eq.~(\ref{mass_matrix}). The mass eigen values are taken to be
$m_{s_1}<m_{s_2}$.
This type of mixing terms are important to induce large contribution to
muon g-2. The contributions to muon g-2 are summarized by
\begin{eqnarray}
a_\mu^{\rm new}&=& -\frac{Q_\chi m_\mu^2}{16\pi^2}
\sum_i \left\{
\left(y_L^2
 |V_{2i}|^2+y_R^2|V_{1i}|^2\right)(C_{11}+C_{21})(s_i,\chi,\chi;p,-q)
\right.\nonumber \\
&&\left.
\hspace{4cm}+2y_L y_R \frac{m_\chi}{m_\mu}{\rm Re}(V_{2i}V_{1i}^*) C_{11}(s_i,\chi,\chi;p,-q)
\right\}
\nonumber\\
&+&\frac{Q_2 m_\mu^2}{16\pi^2}
\sum_i\left\{
\left(
y_L^2 |V_{2i}|^2+y_R^2 |V_{1i}|^2
\right)(C_{12}+C_{22})(s_i,s_i,\chi;q,p-q)
\right.\nonumber \\
&&\left.
\hspace{4cm}+2y_L y_R \frac{m_\chi}{m_\mu}{\rm Re}(V_{2i}V_{1i}^*) C_{12}(s_i,s_i,\chi;q,p-q)
\right\},
\label{muong2_model2}
\end{eqnarray}
where $p$ and $p-q$ are momenta of external muons, and $q$ is a photon 
momentum and a limit $q^2\rightarrow 0$ is taken.
Here we show the explicit expressions of above Passarino-Veltman functions:
\begin{eqnarray}
(C_{11}+C_{21})(s_i,\chi,\chi;p,-q)
&=&\frac{1}{m^2_{s_i}}\frac{2+3y_i-6y_i^2+y_i^3+6y_i\ln
 y_i}{6(1-y_i)^4},
\\
C_{11}(s_i,\chi,\chi;p,-q)
&=&-\frac{1}{m^2_{s_i}}\frac{3-4y_i+y_i^2+2\ln y_i}{2(1-y_i)^3},
\\
(C_{22}+C_{12})(s_i,s_i,\chi;q,p-q)
&=&\frac{1}{m^2_{s_i}}\frac{1-6y_i +3y_i^2 +2y_i^3-6y_i^2\ln
 y_i}{6(1-y_i)^4},
\\
C_{12}(s_i,s_i,\chi;q,p-q)
&=&\frac{1}{m^2_{s_i}}\frac{1-y_i^2+2y_i \ln y_i}{2(1-y_i)^3},
\end{eqnarray}
where $y_i=m^2_\chi/m^2_{s_i}$ and we neglected the higher order of
$O(m^2_\mu/m^2_{s_i})$.

The effective operator which expresses the muon g-2 is written by
\begin{eqnarray}
{\cal L}=\frac{v}{\Lambda^2} \mu_R \sigma^{\mu \nu} \mu_L F_{\mu
 \nu}+{\rm h.c.},
\end{eqnarray}
where $v$ is a vacuum expectation value of Higgs boson, and $F_{\mu
\nu}$ is a field strength of photon field, and $\Lambda$ is a typical scale 
related to new physics. As one see, the chirality of
muon has to flip in this interaction. In the case where only
right-handed muon has the new Yukawa interaction, the chirality flipping
of muon happens in the external muon line in the loop diagram. 
On the other hands, in the
present case where both right- and left-handed muons have the new
Yukawa interactions, the chirality flipping can occur in the internal
fermion line, which is proportional to the mass of the fermion
$\chi$. That is why there are terms which are proportional to 
$y_R y_L m_\chi/m_\mu$ in Eq.~(\ref{muong2_model2}). This contribution
enhances the effects of muon g-2. Therefore, it is very important to
explain the anomaly of muon g-2.

\begin{figure}[ht]
\centerline{
\includegraphics[width=8.5cm,angle=0]{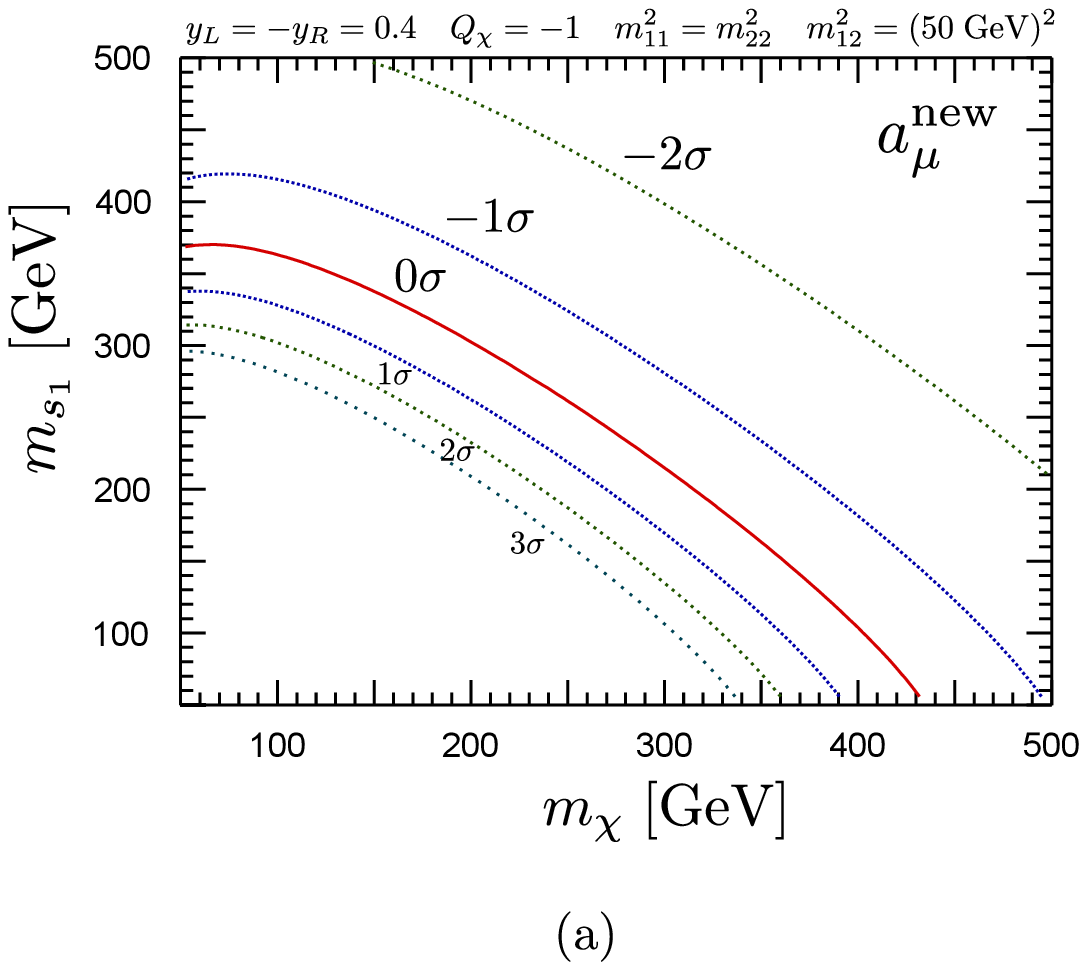}
\includegraphics[width=8.5cm,angle=0]{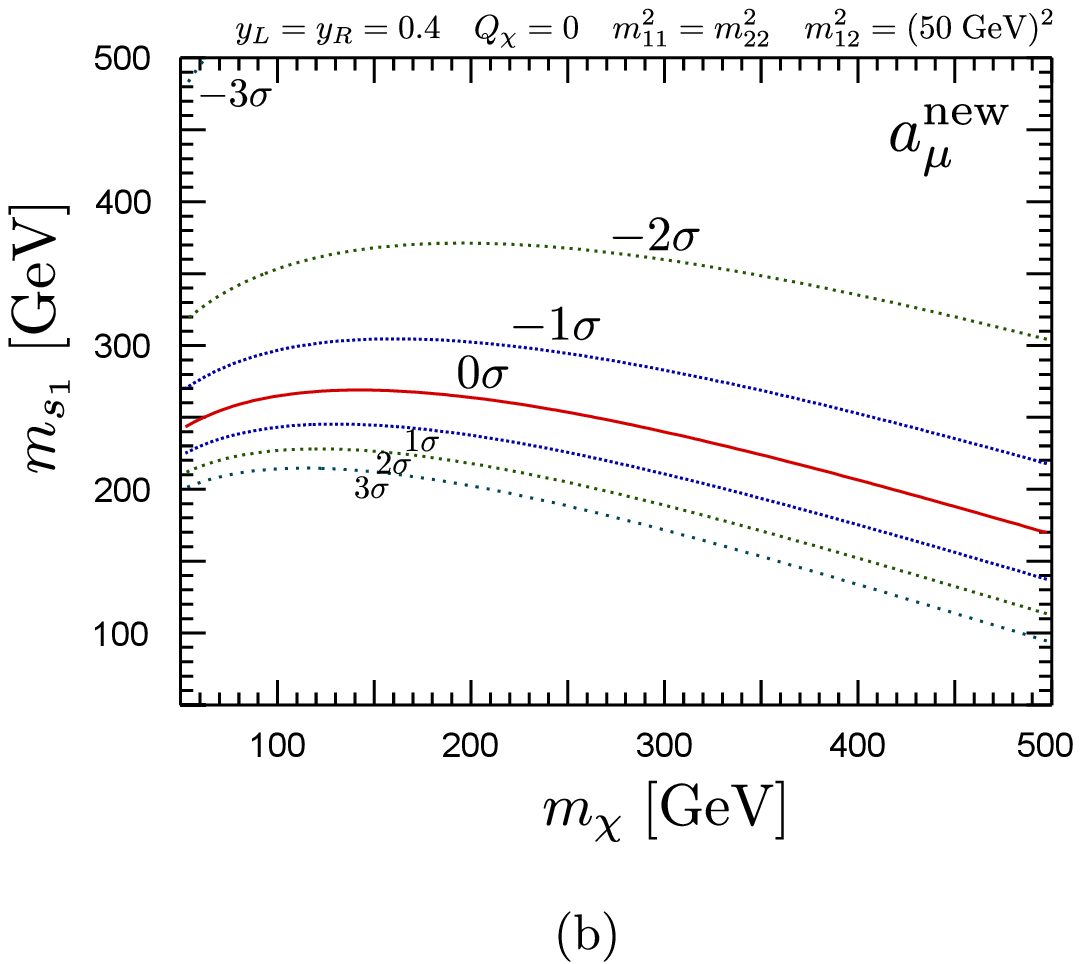}
}
\caption{New physics contribution to muon g-2 ($a_\mu^{\rm new}$)
as a function of $m_\chi$ and $m_{s_1}$ in case of (a) $Q_\chi=-1$ and 
(b) $Q_\chi=0$. 
Contours for $(a_\mu^{\rm new}/10^{-10})=2.1,~10.1,~18.1,~26.1,~34.1,~42.1$ and 
$50.1$ (corresponding to $-3\sigma,~-2\sigma,~-1\sigma,~0\sigma,~1\sigma,~
2\sigma$ and $ 3\sigma$ deviation from the measured value) are shown.
Here we assume that
$m_{11}^2=m_{22}^2$, $m_{12}^2=(50~{\rm GeV})^2$. We also take (a)
$y_L=-y_R=0.4$ and (b) $y_L=y_R=0.4$. 
}
\label{muong2_model2-1}
\end{figure}
In Fig.~\ref{muong2_model2-1}, we show $a_\mu^{\rm new}$ as a function
of $m_\chi$ and $m_{s_1}$ (which is a mass of lighter scalar state) in case of
(a) $Q_{\chi}=-1$ and (b) $Q_\chi=0$. 
Contours for $(a_\mu^{\rm new}/10^{-10})=2.1$, 10.1, 18.1, 26.1, 34.1, 42.1
and 50.1 (corresponding to $-3\sigma,~-2\sigma,~-1\sigma,~0\sigma,~1\sigma,~
2\sigma$ and $ 3\sigma$ deviation from the measured value) are shown.
Here we took $m^2_{11}=m^2_{22}$, $m_{12}^2=(50~{\rm
GeV})^2$, and (a) $y_L=-y_R=0.4$ (b) $y_L=y_R=0.4$. 
Note that in order to generate the positive contribution
to muon g-2 in Fig.~\ref{muong2_model2-1} (a), the sign of new Yukawa
couplings had to be taken as $y_Ly_R<0$.  On the other hand, in
Fig.~\ref{muong2_model2-1} (b), the sign of $y_L y_R$ should be positive.
As shown in Fig.~\ref{muong2_model2-1}, compared to Fig.~\ref{fig_amu_2}, 
even smaller Yukawa couplings and heavier new particles  
can accommodate the anomaly of muon g-2 because of the enhancement mentioned
above. As can be seen from Fig.~\ref{muong2_model2-1}, when $y_L\sim y_R\sim O(1)$,
the EW scale new particles are expected in order to explain the anomaly of muon g-2.
Therefore, we will analyze the effects on the EW precision observables and show
the consistent region of parameters in a next section.

\section{Effects on electroweak observables}

In the previous section, we showed that the relatively light new particles
are required in order to explain the anomaly of muon g-2. In addition,
in the case where the only right-handed muon has new Yukawa interaction,
the new Yukawa coupling should be relatively large for the anomaly of muon g-2.
Therefore, we should check if the relatively light particles and
the relatively large Yukawa coupling are consistent with the EW
precision measurements. 
Although models which we consider in this paper may be only a part of complete models,
we should know how these new particles affect EW observables.

In this section, we adopt the formalism in
Refs.\cite{Cho:2011rk,Cho:1999km,Hagiwara:1994pw} in order to include the oblique corrections as
well as the vertex corrections in the EW observables. First we briefly summarize the
formalism~\cite{Cho:2011rk,Cho:1999km,Hagiwara:1994pw}. 

In the presence of EW to TeV scale physics, it is well known that the
oblique corrections in gauge boson self-energy are important. They
are parameterized by Peskin-Takeuchi's S, T, 
U parameters~\cite{Peskin:1990zt,Altarelli:1990zd,Maksymyk:1993zm}:
\begin{eqnarray}
\frac{\alpha S}{4s_W^2 c_W^2}&=&
\frac{\Pi_{ZZ}(M_Z^2)-\Pi_{ZZ}(0)}{M_Z^2}
-\frac{c_{2W}}{c_W s_W} \frac{\Pi_{Z\gamma}(M_Z^2)}{M_Z^2}
-\frac{\Pi_{\gamma \gamma}(M_Z^2)}{M_Z^2},\\
\alpha T&=& \frac{\Pi_{WW}(0)}{M_W^2}-\frac{\Pi_{ZZ}(0)}{M_Z^2},\\
\frac{\alpha U}{4 s_W^2} &=& \frac{\Pi_{WW}(M_W^2)-\Pi_{WW}(0)}{M_W^2}
-c_W^2\frac{\Pi_{ZZ}(M_Z^2)-\Pi_{ZZ}(0)}{M_Z^2}\nonumber \\
&&-2s_W c_W \frac{\Pi_{Z \gamma}(M_Z^2)}{M_Z^2}
-s_W^2 \frac{\Pi_{\gamma \gamma}(M_Z^2)}{M_Z^2}.
\end{eqnarray}
Here we use the notation $c_W$ and $s_W$ to refer to the cosine and sine of
the weak mixing angle and $c_{2W}=c_W^2-s_W^2 $. 
In addition, in Refs.\cite{Cho:2011rk,Cho:1999km}, $R_Z$ and $R_W$ parameters are
introduced in order to account for the smaller corrections:
\begin{eqnarray}
\frac{\alpha R_Z}{4s_W^2 c_W^2} &=&
 \left. \frac{d\Pi_{ZZ}(p^2)}{dp^2}\right|_{p^2=M_Z^2}-\frac{\Pi_{ZZ}(M_Z^2)-\Pi_{ZZ}(0)}{M_Z^2},\\
\frac{\alpha R_W}{4s_W^2} &=&
 \frac{\Pi_{WW}(M_Z^2)-\Pi_{WW}(M_W^2)}{M_Z^2-M_W^2}-\frac{\Pi_{WW}(M_W^2)-\Pi_{WW}(0)}{M_W^2}.
\end{eqnarray}
Furthermore, new particles have effects on the running QED coupling constant $\alpha(M_Z^2)$~\cite{Hagiwara:2011af,Hagiwara:1994pw}:
\begin{eqnarray}
\alpha(M_Z^2)=\frac{\alpha}{1-\Delta\alpha_{\rm
 lep}(M_Z^2)-\Delta\alpha_{\rm had}^{(5)}(M_Z^2)-\Delta\alpha_{\rm top}(M_Z^2)-\Delta\alpha_{\rm new}(M_Z^2)}
\end{eqnarray}
Here $\Delta\alpha_{\rm lep}(M_Z^2)$, $\Delta\alpha_{\rm had}^{(5)}(M_Z^2)$,
 $\Delta\alpha_{\rm top}(M_Z^2)$, and $\Delta\alpha_{\rm new}(M_Z^2)$ are the leptonic,
 the five-flavor hadronic, the top quark and the new physics
 contributions to the running of the QED coupling constant respectively.
The new physics contributions to the running of the QED coupling constant $\Delta\alpha_{\rm new}(M_Z^2)$ are defined as
\begin{eqnarray}
\Delta\alpha_{\rm
 new}(M_Z^2)=\left. \frac{\Pi_{\gamma\gamma}(M_Z^2)}{M_Z^2}-\frac{\Pi_{\gamma\gamma}(p^2)}{p^2} \right|_{p^2=0}
\end{eqnarray}
These oblique parameters contribute to EW observables, as shown in Refs.
\cite{Cho:2011rk,Cho:1999km,Hagiwara:1994pw}. 

If muon has $O(1)$ new Yukawa couplings,\footnote{In this paper, we assume
that electron and tau do not have $O(1)$ extra Yukawa couplings, for simplicity.}
it induces non-universal vertex corrections
to the $Z\mu^+\mu^-$ coupling.
The standard model coupling of $Z\mu^+\mu^-$ is given by
\begin{eqnarray}
i\frac{g}{c_W}\gamma_\mu [g^{\rm SM,\mu}_L P_L+g^{\rm SM,\mu}_R P_R],
\end{eqnarray}
where $g^{\rm SM,\mu}_{L,R}$ are the standard model couplings, and their
tree level contributions are $g_L^{\rm SM,\mu}= -\frac{1}{2}+s_W^2$ and
$g_R^{\rm SM,\mu}=s_W^2$.
When we take into account of corrections to $Z\mu^+\mu^-$ vertex via the
new Yukawa coupling (including the corresponding wave function
renormalization), the $Z\mu^+\mu^-$ coupling is modified by
\begin{eqnarray}
i\frac{g}{c_W}\gamma_\mu \left[(g^{\rm SM,\mu}_L+\Delta g_L^\mu) P_L+
(g^{\rm SM,\mu}_R+\Delta g_R^\mu) P_R\right].
\end{eqnarray}
Here the vertex corrections generated by new particles are parameterized
by $\Delta g^\mu_{L,R}$. 

The corrections to $W \mu \nu_\mu$ vertex generates the corrections to the
Fermi constant $G_F$ and therefore, we parameterize it as $\Delta \bar{\delta}_G$.
The $\Delta \bar{\delta}_G$ for the new physics contributions
from the vertex and box diagrams to the
$\mu$-decay process is defined as
\begin{eqnarray}
G_F=G_F^{\rm SM + ob. }+ \frac{g^2}{4\sqrt{2}M_W^2}\Delta \bar{\delta}_G.
\end{eqnarray}
Here $G_F^{\rm SM + ob.}$ is the muon decay constant including
effects of SM radiative corrections and new physics oblique corrections
except the vertex and box corrections from new physics.
\begin{table}
\begin{center}
\begin{tabular}{|c|c||c|c||c|c|}
\hline
 & data & SM fit & pull & Sample model& pull\\
\hline
line-shape \& FB asym.:  & & &  & &\\
$\Gamma_Z$(GeV)& 2.4952(23) & 2.4954& -0.1 & 2.4963& -0.5\\
$\sigma_h^0$ (nb) & 41.541(37) &41.479 & 1.7&41.479  & 1.7\\
$R_e$ & 20.804(50) & 20.740& 1.3 & 20.741& 1.3\\
$R_\mu$ & 20.785(33) & 20.740& 1.4 & 20.740& 1.3\\
$R_\tau$ & 20.764(45) & 20.787& -0.5 & 20.788& -0.5\\
$A_{\rm FB}^{0,e}$ & 0.0145(25) & 0.0163& -0.7 & 0.0163& -0.7\\
$A_{\rm FB}^{0,\mu}$ & 0.0169(13) & 0.0163& 0.5 & 0.0163 & 0.4\\
$A_{\rm FB}^{0,\tau}$ & 0.0188(17) & 0.0163& 1.5 & 0.0163& 1.4\\
$\tau$ polarization: & & & & &\\
$A_\tau$ & 0.1439(43) & 0.1472& -0.8 &0.1476 & -0.9\\
$A_e$ & 0.1498(49) & 0.1472& 0.5 & 0.1476& 0.4\\
$b$ and $c$ quark results: & & & & &\\
$R_b$ & 0.21629(66) & 0.21579& 0.8 &0.21580  & 0.7\\
$R_c$ & 0.1721(30)& 0.1723& -0.1 & 0.1722& 0.0\\
$A_{\rm FB}^{0,b}$ & 0.0992(16) &0.1032 &-2.5 &0.1035 &-2.7\\
$A_{\rm FB}^{0,c}$ & 0.0707(35) & 0.0738 & -0.9 & 0.0740& -0.9\\
$A_b$ & 0.923(20) & 0.935  & -0.6 & 0.935& -0.6\\
$A_c$ & 0.670(27) &0.668 & 0.1 & 0.668& 0.1\\
SLD results: & & & & &\\
$A_e$ & 0.1516(21) &0.1472 &2.1  &0.1476 &1.9\\
$A_\mu$ & 0.142(15) &0.1472 &-0.4 &0.1476 &-0.4 \\
$A_\tau$ & 0.136(15) &0.1472 &-0.8 &0.1476 &-0.8 \\
W mass and width: & & & & & \\
$M_W$ (GeV) & 80.385(15)\cite{TevatronElectroweakWorkingGroup:2012gb} & 80.363& 1.5 & 80.376& 0.6\\
$\Gamma_W$ (GeV) & 2.085(42)& 2.091& -0.1 & 2.092&-0.2\\
muon g-2: & & & & & \\
$a_\mu^{\rm new}(10^{-9})$ & 2.61(0.80) & 0 & 3.3 &3.15 &-0.7\\
\hline
Input parameters & & & & &\\
$\Delta \alpha_{\rm had}^{(5)}(M_Z^2)$ & 0.027626(138) & 0.027592&0.3 &0.027626 &0.0\\
$\alpha_s(M_Z)$ & 0.1184(7) & 0.1184&0.0 & 0.1184& 0.0\\
$m_t$ (GeV) & 173.2(0.9)\cite{Lancaster:2011wr} &173.7 &-0.6 &173.3 &-0.1\\
$m_h$ (GeV) &  &125 & & 125&\\
\hline
$y_L=y_R$,~ $Q_\chi$&- &- &- &0.4,~0 & \\
$m_{\phi_1}$, $m_\chi$ (GeV) & -&- &- &300,~200 & \\
$m_{11}^2=m_{22}^2$, $m_{12}^2$ $({\rm GeV})^2$ &- &- &- &$(250)^2$, $(50)^2$ & \\
\hline
$\chi^2/(d.o.f) $ & & 34.8/(22)& &22.5/(15)&
\\
\hline
\end{tabular}
\end{center}
\caption{Experimental data and theoretical predictions of electroweak observables.
The predicted values of the SM and sample model discussed in Sec. 3.2
 are shown.}
\label{EWObs}
\end{table}
\begin{figure}[t]
\centerline{
\includegraphics[width=8.0cm]{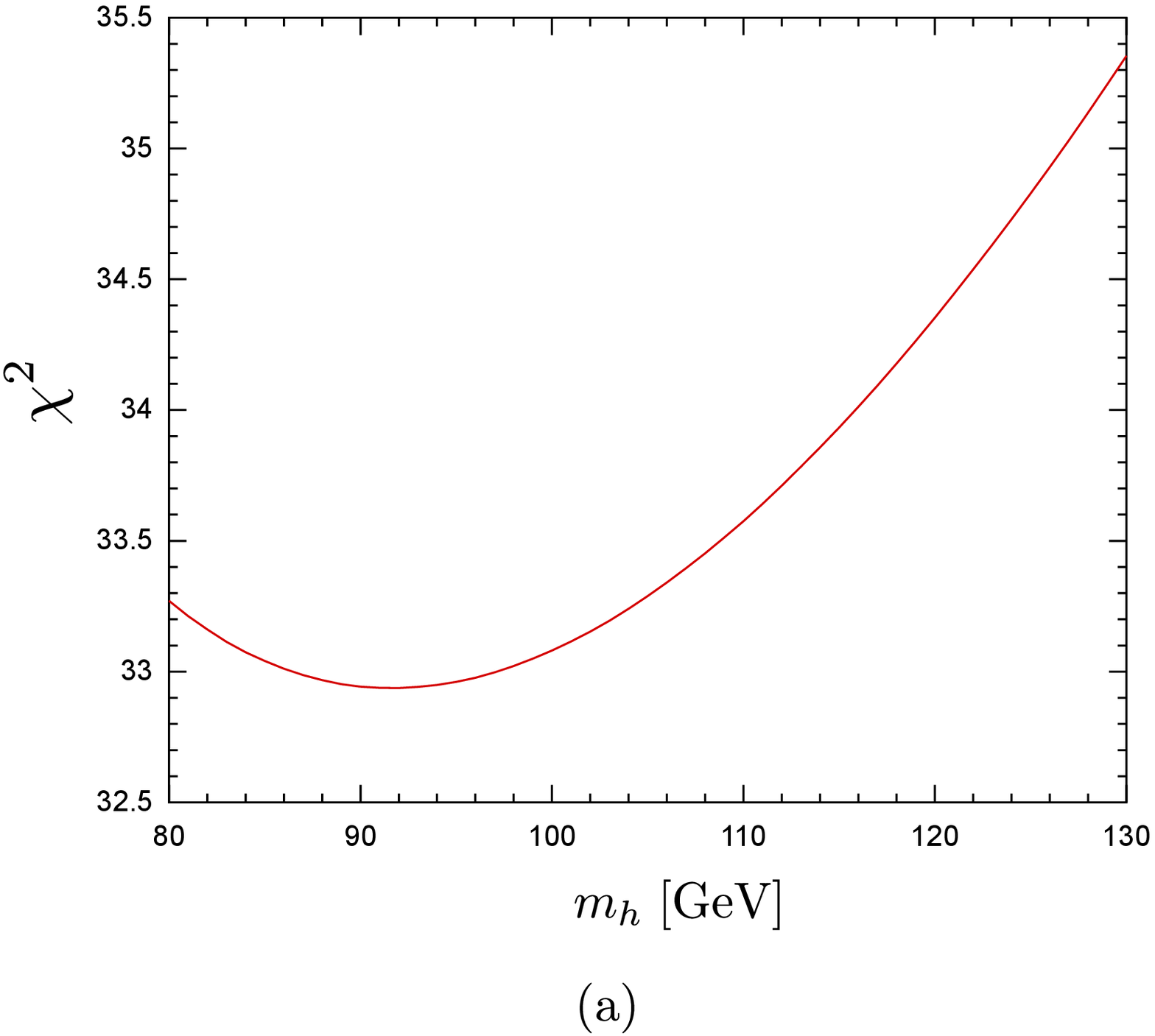}
\includegraphics[width=8.0cm]{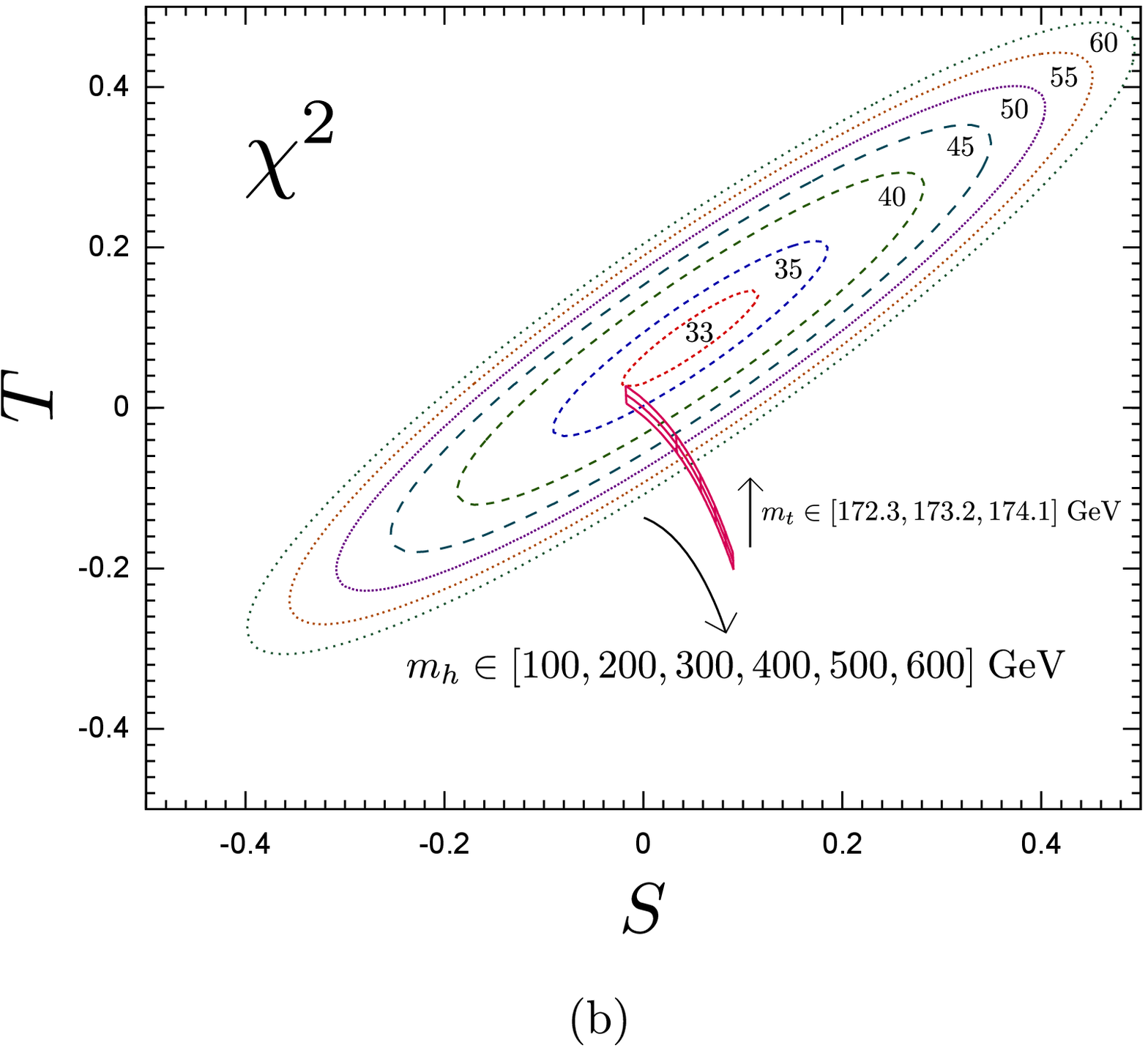}
}
\caption{(a) $\chi^2$ of the EW observables listed in
 Table.~\ref{EWObs} as a function of the Higgs boson mass $m_h$ within the
 SM. (b) $\chi^2$ contours shown in dashed lines in S-T plane, 
assuming that other oblique corrections
($U$, $R_W$, $R_Z$) and vertex corrections are zero. Reference Higgs boson mass and top quark mass
are taken to be $125$ GeV and 173.2 GeV, respectively. We also show the predicted S-T values (shown in solid lines) 
in the SM, varying the Higgs boson mass and top quark mass from the reference values.
}
\label{SMfit}
\end{figure}

Using the formalism in Ref.\cite{Cho:2011rk,Cho:1999km}, one can calculate EW
observables listed in Table~\ref{EWObs} without assuming lepton
universality. 
First, we show a result of the SM fit of the EW observables in
Fig.~\ref{SMfit} (a) as a function of Higgs boson mass $m_h$.
The best fit point is at $m_h=91$ GeV. This is consistent with the
result shown by the LEP electroweak working group~\cite{LEPEW}.
In Table~\ref{EWObs}, we show the SM fit when
Higgs boson mass is assumed to be 125 GeV, which has been suggested by the
latest LHC data. In Fig.~\ref{SMfit}(b), $\chi^2$ contours 
are shown 
in dashed lines in S-T plane. Here we assume that other oblique corrections ($U$, $R_W$, $R_Z$) and vertex corrections
are zero. Reference Higgs boson mass and top quark mass are taken to be 125 GeV, 173.2 GeV, respectively.
The predicted values of S-T parameters are also shown by varying the Higgs boson mass and top quark mass from
the reference values.
We note that the SM fit with lighter Higgs boson is good except for the muon g-2, and 
small positive S and T $(S\sim 0.05,~T\sim 0.1$) can decrease the $\chi^2$ further.
Therefore, we need new physics which largely contributes to muon
g-2, but whose effects on other EW observables are small. In this
section, we analyze the EW observables in models discussed in the previous section.
In this analysis, we assume $m_h=125$ GeV.

\subsection{ Model with SU(2) singlet scalar ($\phi$)
 and singlet Dirac fermion ($\chi$)}

Since the new scalar $\phi$ and new fermion $\chi$ are SU(2) singlet,
they do not couple to W boson. But if it has a QED charge, it can couple to
photon and Z boson. Here we list the corrections to self-energy
functions of gauge bosons in this model.

The singlet fermion ($\chi$) contributions are given by
\begin{eqnarray}
\Pi_{WW}^{(\chi)}(p^2) &=&0 \nonumber \\ 
\Pi_{ZZ}^{(\chi)}(p^2)&=&-\frac{g^2Q_\chi^2s_W^4}{4\pi^2c_W^2}
\left[
m_\chi^2 B_0(\chi,\chi)-p^2\left\{
B_1(\chi,\chi)+B_{21}(\chi,\chi)
\right\}-2(1-\epsilon)B_{22}(\chi,\chi)
\right], \nonumber\\
\Pi_{\gamma \gamma}^{(\chi)}(p^2)&=&
-\frac{e^2}{4\pi^2}Q_\chi^2\left[
m^2_{\chi}B_0(\chi,\chi)-p^2\left\{B_1(\chi,\chi)
+B_{21}(\chi,\chi)\right\}
-2(1-\epsilon)B_{22}(\chi,\chi)
\right], \nonumber \\
\Pi_{\gamma Z}^{(\chi)}(p^2) &=&\frac{geQ_\chi^2s_W^2}{4\pi^2c_W}
\left[
m_\chi^2B_0(\chi,\chi)-p^2\left\{
B_1(\chi,\chi)+B_{21}(\chi,\chi)
\right\}
-2(1-\epsilon)B_{22}(\chi,\chi)
\right]. \nonumber\\
\label{singlet_chi}
\end{eqnarray}
Here $B_X(i,j)=B_X(m_i^2,m_j^2;p)~(X=0,~1,~21,~22)$, 
which are Passarino-Veltman functions and are shown explicitly in Appendix. 
We use the dimensional regularization in space-time dimension $D=4-2\epsilon$.

The singlet scalar ($\phi$) contributions are given by
\begin{eqnarray}
\Pi_{WW}^{(\phi)}(p^2)&=&0
\nonumber
\\
\Pi_{ZZ}^{(\phi)}(p^2)&=&\frac{g^2}{4\pi^2c_W^2}Q_\phi^2 s_W^4
\left\{
B_{22}(\phi,\phi)-\frac{1}{2}A(\phi)
\right\},
\nonumber \\
\Pi_{\gamma \gamma}^{(\phi)}(p^2)&=&
\frac{e^2}{4\pi^2}
Q_\phi^2\left\{B_{22}(\phi,\phi)-\frac{1}{2} A(\phi)\right\},
\nonumber
\\
\Pi_{\gamma Z}^{(\phi)}(p^2) &=&
-\frac{ge}{4\pi^2c_W}
Q_\phi^2 s_W^2\left\{
B_{22}(\phi,\phi)-\frac{1}{2}A(\phi)
\right\}.
\end{eqnarray}
Here Passarino-Veltman functions $B_{22}(i,j)(=B_{22}(m_i^2,m_j^2;p))$ and $A(\phi)$ are shown in Appendix. 
We can easily show that Peskin-Takeuchi STU parameters in this case
vanish ($S=T=U=0$), since the new particles do not have SU(2)$_L$ interactions.
Thus, the leading contributions to the oblique corrections
are $R_Z$ parameter and $\Delta\alpha_{\rm new}(M_Z^2)$:
\begin{eqnarray}
R_Z&=&\frac{4s_W^4 Q_\chi^2}{3\pi}\left[1+\frac{6m_\chi^2}{M_Z^2}
\left\{1-\frac{\frac{4m_\chi^2}{M_Z^2}}{\sqrt{\frac{4m_\chi^2}{M_Z^2}-1}}
\tan^{-1}\left[\frac{1}{\sqrt{\frac{4m_\chi^2}{M_Z^2}-1}}\right]\right\}\right]
\nonumber \\
&&+\frac{s_W^4 Q_\phi^2}{3\pi}\left[1-\frac{12m_\phi^2}{M_Z^2}
\left\{1-\sqrt{\frac{4m_\phi^2}{M_Z^2}-1}
\tan^{-1}\left[\frac{1}{\sqrt{\frac{4m_\phi^2}{M_Z^2}-1}}\right]\right\}\right],
\end{eqnarray}

\begin{eqnarray}
\Delta\alpha_{\rm new}(M_Z^2)&=&-\frac{5\alpha Q_\chi^2}{9\pi}\left[
1+\frac{12 m_\chi^2}{5M_Z^2}-\frac{6}{5}
\left(1+\frac{2m_\chi^2}{M_Z^2}\right)
\sqrt{\frac{4m_\chi^2}{M_Z^2}-1}
\tan^{-1}\left[\frac{1}{\sqrt{\frac{4m_\chi^2}{M_Z^2}-1}}\right]\right]
\nonumber \\
&&-\frac{2\alpha Q_\phi^2}{9\pi}\left[
1-\frac{3m_\phi^2}{M_Z^2}
+\frac{3}{4}\sqrt{\frac{4m_\phi^2}{M_Z^2}-1}
\tan^{-1}\left[\frac{1}{\sqrt{\frac{4m_\phi^2}{M_Z^2}-1}}\right]\right],
\end{eqnarray}
where we assume $2m_\chi>M_Z$ and $2m_\phi>M_Z$.

\begin{figure}[ht]
\centerline{
\includegraphics[width=15.0cm,angle=0]{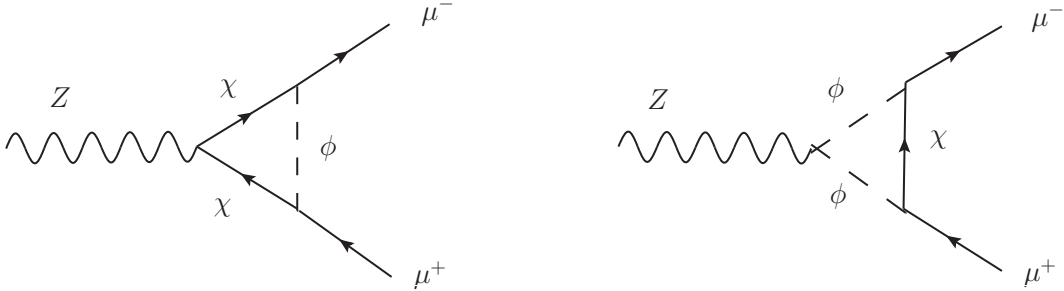}
}
\caption{Feynman diagrams for vertex corrections to $Z\mu^+\mu^-$ coupling.}
\label{vertex_fig}
\end{figure}

As discussed in the previous section, in order to explain the anomaly of
muon g-2, the new Yukawa coupling in this model should be relatively large.
In this case, the potentially large vertex corrections may be expected, as
shown in Fig.\ref{vertex_fig}.
The results are expressed by
\begin{eqnarray}
\Delta g_L^\mu&=&0,\\
\Delta g_R^\mu &=&
\frac{y_N^2}{16\pi^2}\left[
-2Q_\phi s_W^2 C_{24}(\phi,\chi,\phi;p,q-p)
\right.\nonumber\\
&&+Q_\chi s_W^2\left\{
\frac{1}{2}-2 C_{24}-M_Z^2(C_{12}+C_{23})+m_\chi^2 C_0
\right\}(\chi,\phi,\chi;q-p,p) \nonumber
\\
&&\left.
-s_W^2(B_0+B_1)(\phi,\chi;p)
\right],
\end{eqnarray}
where $p$, $q-p$ and $q$ are muon, anti-muon and Z-boson momenta, respectively.
Here Passarino-Veltman functions $C_X~(X=0,~12,~23,~24)$ and $B_X~(X=0,~1)$ are explicitly
shown in Appendix.
\begin{figure}[ht]
\centerline{
\includegraphics[width=8cm]{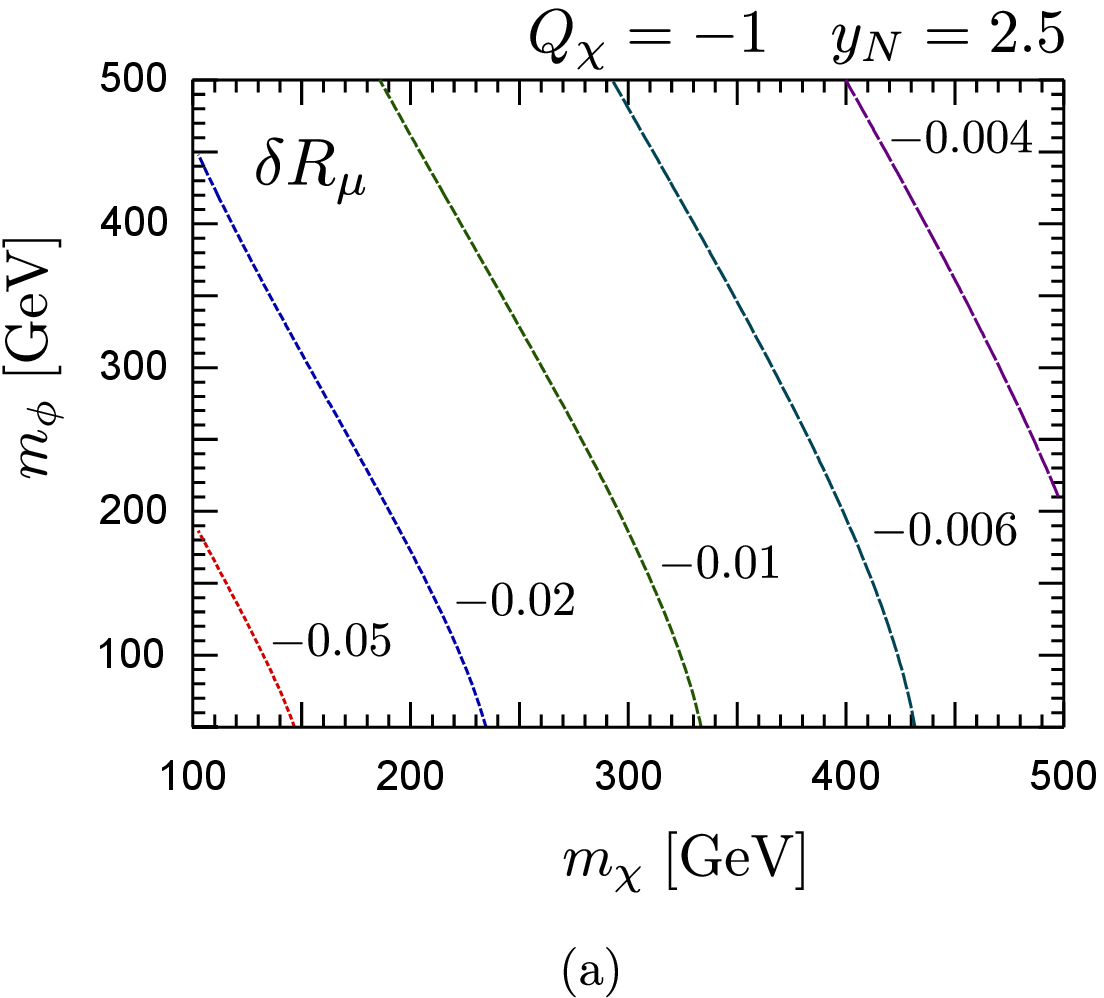}
\includegraphics[width=8cm]{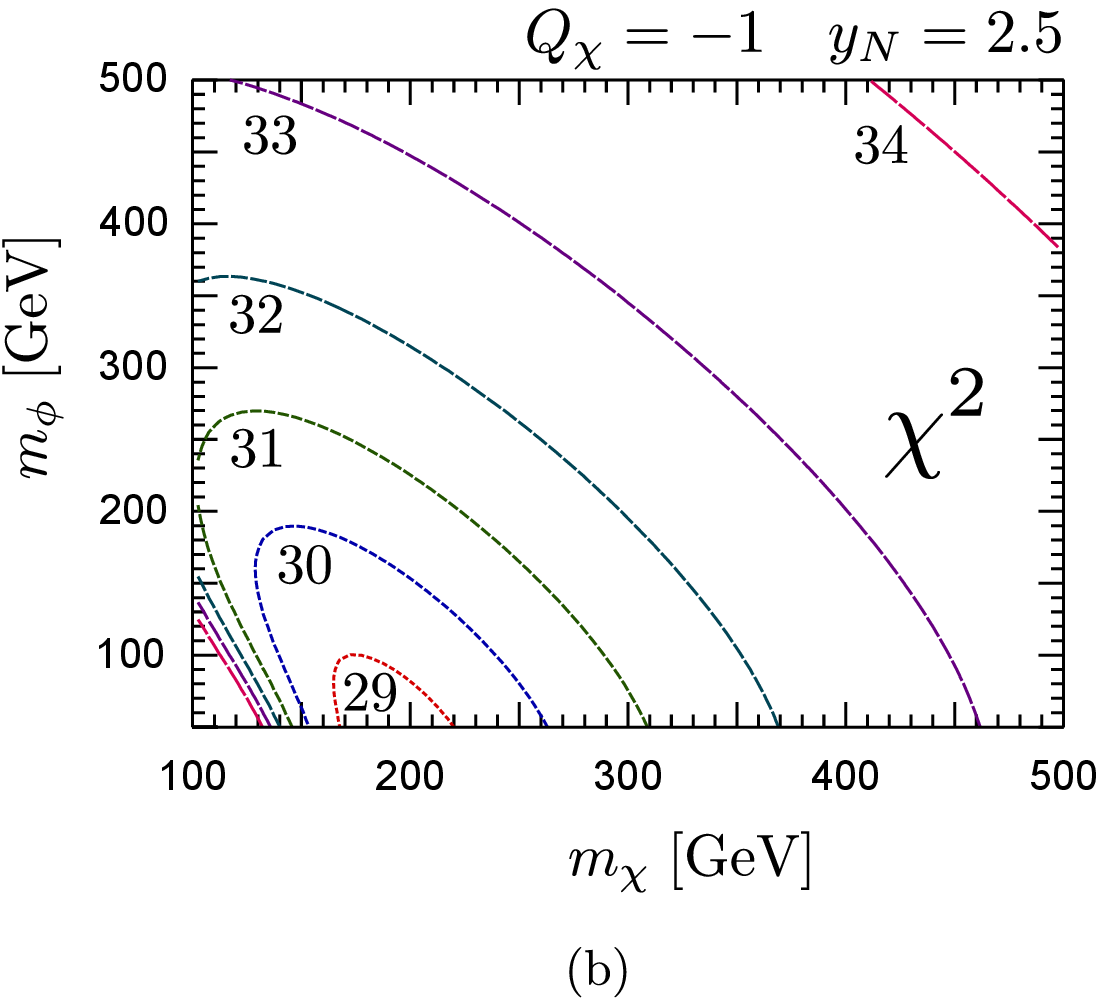}
}
\caption{(a) Effect of vertex corrections on $R_\mu$. $\delta
 R_\mu=R_\mu-R_\mu'$ are shown as a function of $m_\chi$ and $m_\phi$ in
 case of $Q_\chi=-1$ and $y_N=2.5$, where $R_\mu$ includes all corrections, and $R_\mu'$
contains all corrections except for the vertex corrections. (b) $\chi^2$ as a function of
 $m_\chi$ and $m_\phi$ in case of $Q_\chi=-1$ and $y_N=2.5$.}
\label{vertex}
\end{figure}
\begin{figure}[ht]
\centerline{
\includegraphics[width=8cm]{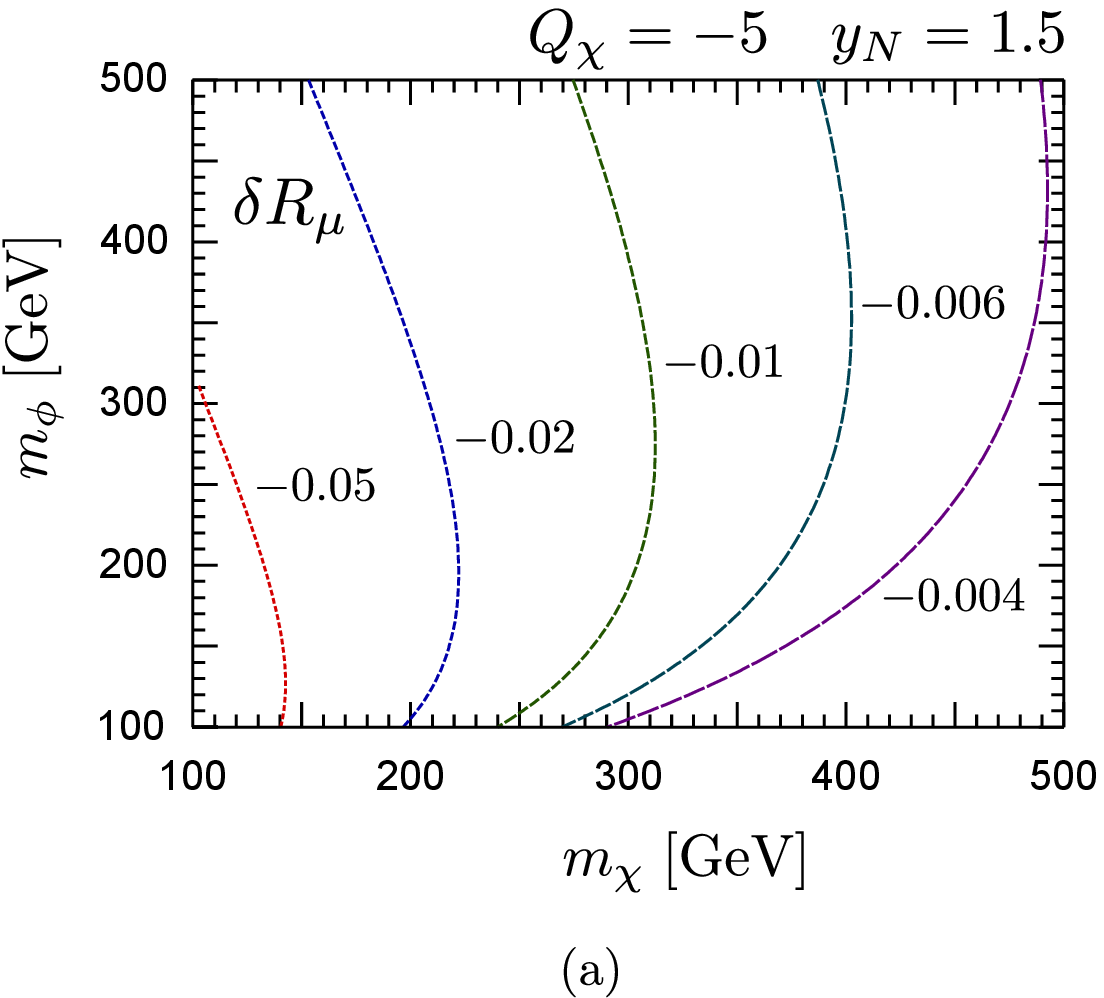}
\includegraphics[width=8cm]{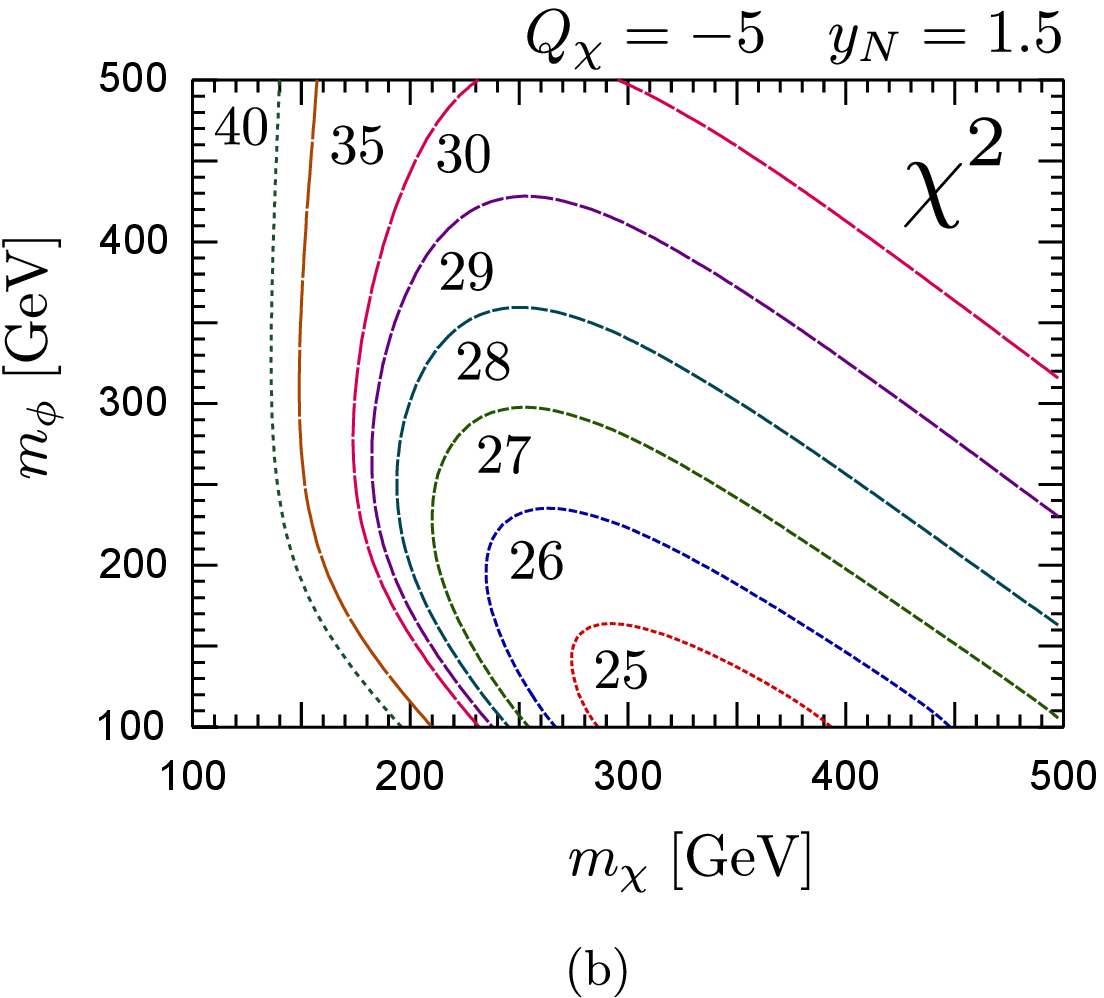}
}
\caption{(a) Effect of vertex corrections on $R_\mu$. $\delta
 R_\mu=R_\mu-R_\mu'$ is shown as a function of $m_\chi$ and $m_\phi$ in
 case of $Q_\chi=-5$ and $y_N=1.5$, similar to Fig.~\ref{vertex}(a).
(b) $\chi^2$ as a function of
 $m_\chi$ and $m_\phi$ in case of $Q_\chi=-5$ and $y_N=1.5$. }
\label{vertex2}
\end{figure}

The vertex corrections affect EW observables such as $R_\mu$, $A^{0,\mu}_{\rm FB}$, and 
$A_\mu$.  Among these observables, we find that the effect on the $R_\mu$ is the most important in the fit 
because $R_\mu$ is most precisely measured.
In Fig.\ref{vertex} (a), we show the effect of vertex corrections in
$R_\mu$, that is, $\delta R_\mu=R_\mu-R_\mu'$ as a function of $m_\chi$ and $m_\phi$ in case of 
$Q_\chi=-1$ and $y_N=2.5$. Here $R_\mu$ is a theoretical prediction
in this model containing all corrections, on the other hand, $R_\mu'$ contains all corrections
except for the vertex corrections. The difference $\delta R_\mu$ shows the effect of the
vertex corrections. Note that the size of experimental error of $R_\mu$
is $0.033$ for $1\sigma$, as shown in Table.~\ref{EWObs}.
As one can see in Fig.~\ref{vertex}(a), the vertex
correction can change the prediction of $R_\mu$ about $1\sigma$ in
the region where the muon g-2 can be explained by the new physics contributions.
Since the SM prediction of $R_\mu$ is smaller than that of the experimental result (shown in Table.~\ref{EWObs}),
the vertex corrections do not help to improve the prediction of $R_\mu$. Therefore, the region of small $\chi$ mass
is very constrained.

In Fig.~\ref{vertex} (b), $\chi^2$ is shown as a function of $m_\chi$
and $m_\phi$ in case of $Q_\chi=-1$ and $y_N=2.5$. As one can see from the figure,
$\chi^2$ is large in the region of small $m_\chi$ because of the large vertex corrections, discussed above.
In the region of larger $m_\chi$ and $m_\phi$, it is difficult to explain the anomaly of muon g-2 because
new particles are too heavy, and hence the $\chi^2$ gets larger. The minimum of $\chi^2$ is around $m_\chi\sim 200$ GeV
and $m_\phi\sim 100$ GeV, and we see that the light new particles are favored by the EW observables including muon g-2.

In Fig.~\ref{vertex2} (a),  the effect of vertex corrections on $R_\mu$,
$\delta R_\mu$ is shown in case of $Q_\chi=-5$ and $y_N=1.5$, similar to Fig.~\ref{vertex} (a).
The vertex corrections can be as large as $1\sigma$ of the experimental error of $R_\mu$, and they increase the $\chi^2$.
In Fig.~\ref{vertex2} (b), the $\chi^2$ is shown as a function of $m_\chi$ and $m_\phi$. As one can see, the region of
small $m_\chi$ is disfavored by the vertex corrections of $R_\mu$, as shown in Fig.~\ref{vertex2} (a). The minimum of
the $\chi^2$ is around $m_\chi\sim 300$ GeV and $m_\phi\sim 100$ GeV. Therefore, the EW precision measurements
(including muon g-2) prefer the relatively light new particles in this case.

\subsection{Model with SU(2) doublet ($\Phi$) and
singlet ($\phi$) scalar, and singlet fermion ($\chi$)}

The singlet fermion contributions are same as those in the previous case.
Here, we list the one loop contributions to gauge boson self-energies induced by scalar sectors.
\begin{eqnarray}
\Pi_{WW}^{(s)}(p^2)&=& \frac{g^2}{8\pi^2}\left[
\sum_i\left\{
|V_{2i}|^2 B_{22}(\phi_1,s_i)-\frac{1}{4}|V_{2i}|^2 A({s_i})
\right\}-\frac{1}{4}A(\phi_1)
\right],
\\
\Pi_{ZZ}^{(s)}(p^2)&=&\frac{1}{4\pi}\frac{g^2}{c_W^2}\left[
\left(
\frac{1}{2}-Q_1 s_W^2
\right)^2
\left\{
B_{22}(\phi_1,\phi_1)-\frac{1}{2} A(\phi_1)
\right\}
\right.
\nonumber \\
&&+\sum_{ij}\left\{
\left(
-\frac{1}{2}-Q_2s_W^2
\right)^2|V_{2i}|^2|V_{2j}|^2
+Q_2^2 s_W^4 |V_{1i}|^2|V_{1j}|^2
\right.
\nonumber \\
&&-\left.
Q_2s_W^2 \left(
-\frac{1}{2}-Q_2 s_W^2
\right)
(V_{2i}^*V_{1i}V_{1j}^* V_{2j}+V_{2j}^*V_{1j}V_{1i}^*V_{2i})
\right\}B_{22}(s_i,s_j)
\nonumber \\
&&\left.-\sum_i\left\{
\left(-\frac{1}{2}-Q_2s_W^2
\right)^2 |V_{2i}|^2+Q_2^2 s_W^4 |V_{1i}|^2
\right\}\frac{1}{2}A(s_i)
\right],
\end{eqnarray}
\begin{eqnarray}
\Pi_{\gamma \gamma}^{(s)}(p^2)&=&
\frac{e^2}{4\pi^2}\left[
Q_1^2 \left\{
B_{22}(\phi_1,\phi_1)-\frac{1}{2}A(\phi_1)
\right\}
\right.\nonumber \\
&&~~~~~~~~~~~~~~~~~~~~~~~~~+\left.Q_2^2\sum_i
\left\{
B_{22}(s_i,s_i)-\frac{1}{2}A(s_i)\right\}
\right],
\\
\Pi_{\gamma Z}^{(s)}(p^2) &=&
\frac{e}{4\pi^2}\frac{g}{c_W}
\left[
Q_1 \left(
\frac{1}{2}-Q_1 s_W^2
\right)\left\{
B_{22}(\phi_1,\phi_1)-\frac{1}{2}A(\phi_1)
\right\}
\right.
\nonumber \\
&&
+Q_2\sum_i\left\{
\left(
-\frac{1}{2}-Q_2 s_W^2
\right)|V_{2i}|^2-Q_2 s_W^2 |V_{1i}|^2
\right\}
\nonumber \\
&&\left. ~~~~~~~~~~~~~~~~~~~~~~~~~~\times\left\{
B_{22}(s_i,s_i)-\frac{1}{2}A(s_i)
\right\}
\right],
\end{eqnarray}
where Passarino-Veltman functions $B_X(i,j)(=B_X(m_i^2,m_j^2;p))$ and $A(i)$ are shown in Appendix.
Unlike the previous model, SU(2) scalar doublet can contribute to
the STU parameters. Therefore, the dominant quantum corrections are
represented by S and T parameters. To understand a behavior of S and T
parameters, here we show the approximate expression of S and T parameters,
\begin{eqnarray}
S&\simeq& \frac{Y_{\Phi}}{6\pi}\Delta+\cdots,\nonumber \\
T&\simeq& \frac{m_{\phi_1}^2}{16 \pi s_W^2 M_W^2} (\Delta)^2+\cdots.
\label{ST}
\end{eqnarray}
Here, for simplicity, we assume that $V_{22}=V_{11}=1,~V_{12}=V_{21}=0$ 
so that $s_2$ state
originates from $\phi_2$ component of SU(2) scalar doublet $\Phi$, 
and $\Delta=(m_{\phi_2}^2-m_{\phi_1}^2)/m_{\phi_1}^2$ where $m_{\phi_2}=m_{s_2}$ 
in this case. Here we ignore the higher order terms of $\Delta$ and $m_Z^2/m_{\phi_1}^2$ 
in Eqs.~(\ref{ST}).
Note that $\Delta$ parameterizes non-degeneracy in the SU(2) scalar doublet $\Phi$.
$Y_{\Phi}$ is a hypercharge of $\Phi$, 
$Y_{\Phi}=\frac{1}{2}+Q_2$. We note that the non-degeneracy of SU(2) scalar doublet 
generates non-zero T and non-zero S, and T is always positive, but
S can be positive or negative depending on signs of $Y_\Phi$ and $\Delta$ in this model.

\begin{figure}[ht]
\centerline{
\includegraphics[width=13.0cm]{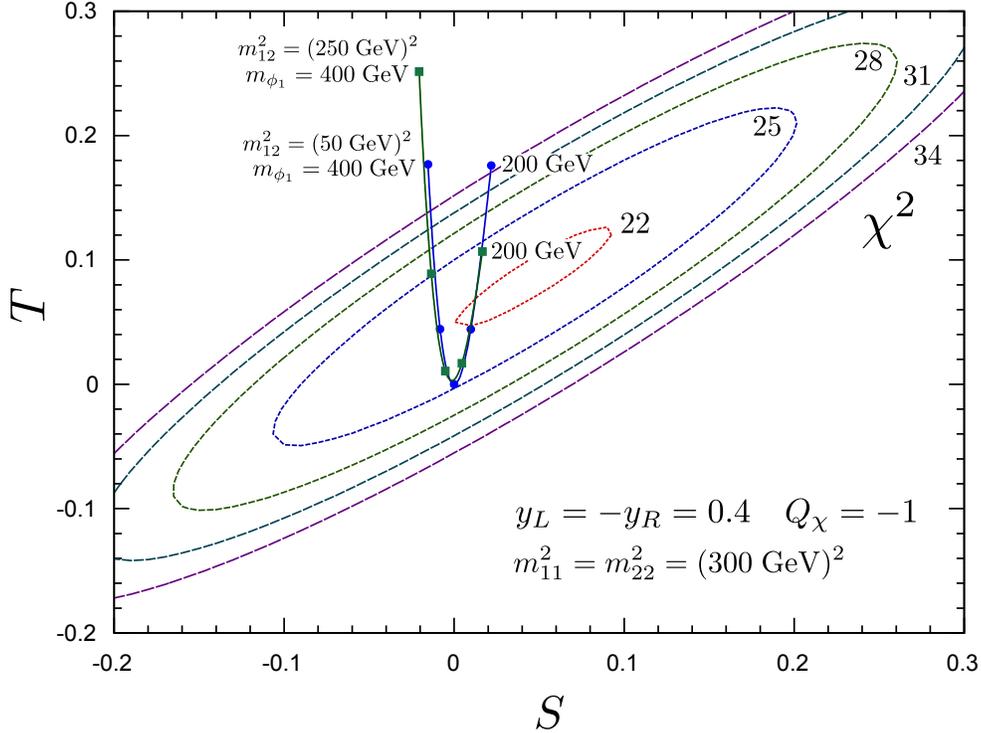}
}
\caption{$\chi^2$ contours (without including muon g-2 result) in S-T plane (
$\chi^2=22$ to 34 in dashed lines) and 
S-T values (solid lines with points) in this model. Here we take 
$y_L=-y_R=0.4$, $Q_\chi=-1$ (corresponding to $Q_1=1$ and $Q_2=0$), and
$m_{11}^2=m_{22}^2=(300~{\rm GeV})^2$.
Two solid lines correspond to two different values of $m_{12}^2$, $m_{12}^2=(50~{\rm GeV})^2$ 
and $(250~{\rm GeV})^2$, varying $m_{\phi_1}$ from 200 GeV to 400 GeV at 50 GeV step.
}
\label{ST_para1}
\end{figure}

In Fig.~\ref{ST_para1}, we show numerical result of $\chi^2$ 
contours in S-T plane (without including muon g-2 result), 
shown in dashed lines. To draw the $\chi^2$ contours in S-T plane, we assume that other oblique corrections 
($U$, $R_Z$, and $R_W$) and vertex corrections are zero.
As one can see from the figure, slightly positive S and T ($S\sim 0.05$ and $T\sim 0.1$) are favored by EW observables.
In Fig.~\ref{ST_para1}, we also show the predicted values of S-T parameters in this model, shown in solid lines.
Here we took $y_L=-y_R=0.4$, $Q_\chi=-1$ (corresponding to $Q_1=1$ and $Q_2=0$), and $m_{11}^2=m_{22}^2=(300~{\rm GeV})^2$. 
Two solid lines correspond to two different values of $m_{12}^2$, $m_{12}^2=(50~{\rm GeV})^2$ and $(250~{\rm GeV})^2$.
The points on the solid lines represent the predicted values of S and T, varying $m_{\phi_1}$ from 200 GeV to 400 GeV at 50 GeV step.

When $m_{11}^2=m_{22}^2=(300~{\rm GeV})^2$ and $m_{12}^2=(50~{\rm GeV})^2$,
$m_{s_1}\simeq 296$ GeV and $m_{s_2}\simeq 304$ GeV.
As roughly shown in Eqs.~(\ref{ST}), when $m_{\phi_1}=200$ GeV and $Q_2=0$, $\Delta>0$ and 
$Y_{\Phi}=\frac{1}{2}>0$, and hence $S>0$ and $T>0$. As $m_{\phi_1}$ becomes larger and about
300 GeV, both $S$ and $T$ get closer to zero because the doublet scalars are almost degenerate. 
Then when $m_{\phi_1}$ is increased further,
$S$ gets negative because $\Delta<0$, but $T>0$. In Fig.~\ref{ST_para1}, one can see this behavior. 
Even when we increase $m_{12}^2$,
we can see that this behavior is almost same.
\begin{figure}[ht]
\centerline{
\includegraphics[width=13.0cm]{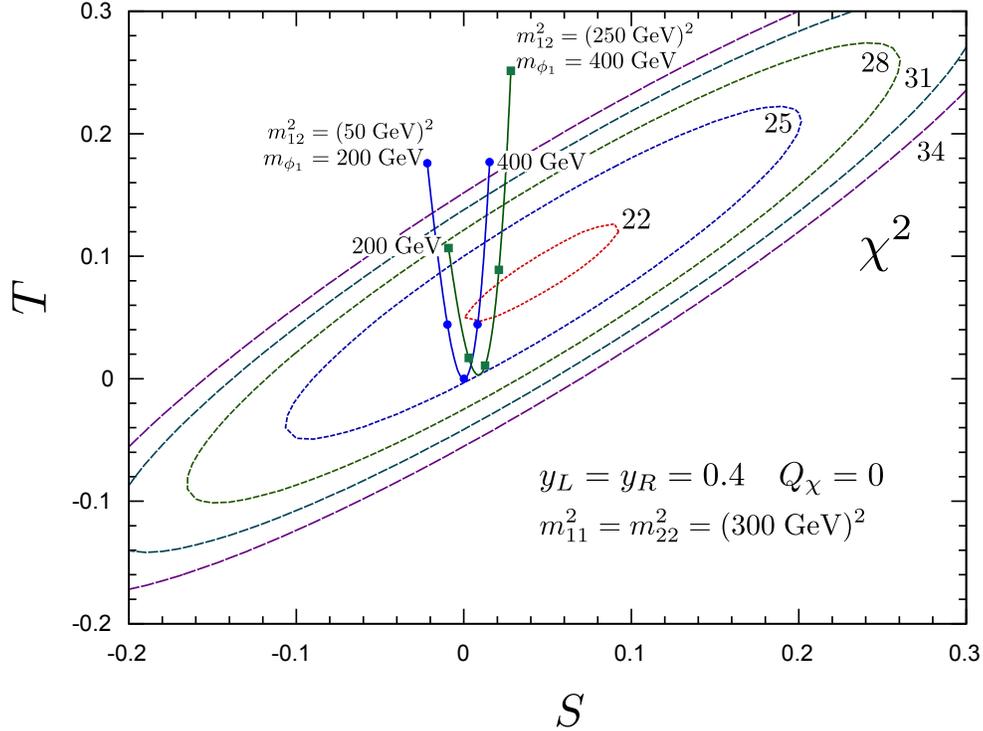}
}
\caption{
Same as Fig.~\ref{ST_para1} except for $y_L=y_R=0.4$ and $Q_\chi=0$ (corresponding to $Q_1=0$ and $Q_2=-1$).
}
\label{ST_para2}
\end{figure}

In Fig.~\ref{ST_para2}, the same figure is shown except for $y_L=y_R=0.4$ and $Q_\chi=0$ (corresponding to
$Q_1=0$ and $Q_2=-1$).
When $Q_\chi=0$ ($Q_2=-1$) and $m_{\phi_1}=200$ GeV, $Y_\Phi=-\frac{1}{2}<0$ and $\Delta>0$, so that $S$
is negative. As $m_{\phi_1}$ gets larger, $S$ gets larger. For $m_{\phi_1}\sim 300$ GeV,
$S\sim 0$ because the doublet scalars are almost degenerate. For $m_{\phi_1}>300$ GeV, $S$ becomes positive. This behavior is 
different from the previous case with $Q_2=0$ since the sign of $Y_\Phi$ is different.
Note that $T$ is always positive
unless SU(2) scalar doublet is degenerate.
As one can see in both cases, the small non-degeneracy of SU(2) scalar doublet can make $\chi^2$ better.

In our numerical analysis, we also include the vertex
corrections in $Z\mu^+\mu^-$, $Z\nu_{\mu}\bar{\nu}_\mu$ and $W\mu\nu_\mu$ vertices even though the
Yukawa couplings $y_L$ and $y_R$ can be small in this model.
The expression of the vertex corrections is summarized as
follows: 
\begin{eqnarray}
\Delta g_L^\mu &=&\frac{y_L^2}{16\pi^2}\left[
2\sum_{ij} \left\{\left(-\frac{1}{2}-Q_\phi s_W^2\right)V_{2i}^* V_{2j}-Q_\phi
   s_W^2 V_{1i}^* V_{1j}\right\} V_{2i}V_{2j}^* C_{24}(s_i,\chi,s_j;
p,q-p)\right.\nonumber \\
&&+\sum_i Q_\chi s_W^2
|V_{2i}|^2 \left\{
\frac{1}{2}-2C_{24}-M_Z^2 (C_{12}+C_{23})+m_\chi^2 C_0\right\}(\chi,s_i,\chi;q-p,p)
 \nonumber \\
&&\left. -\sum_i \left(
-\frac{1}{2}+s_W^2
\right)|V_{2i}|^2(B_0+B_1)(s_i,\chi;p)
\right],\\
\Delta g_R^\mu &=& \frac{y_R^2}{16\pi^2}\left[
2 \sum_{ij} \left\{\left(-\frac{1}{2}-Q_\phi s_W^2\right)V_{2i}^* V_{2j}-Q_\phi
   s_W^2 V_{1i}^* V_{1j}\right\} V_{1i}V_{1j}^* C_{24}(s_i,\chi,s_j;
p,q-p)\right.\nonumber \\
&&+\sum_i Q_\chi s_W^2
|V_{1i}|^2 \left\{
\frac{1}{2}-2C_{24}-M_Z^2 (C_{12}+C_{23})+m_\chi^2 C_0\right\}(\chi,s_i,\chi;q-p,p)
 \nonumber \\
&&\left. -\sum_i s_W^2|V_{1i}|^2(B_0+B_1)(s_i,\chi;p)
\right],\\
\Delta g_L^{\nu_\mu} &=&
\frac{y_N^2}{16\pi^2}\left[
2\left(\frac{1}{2} -s_W^2 Q_{\phi_1} \right) C_{24}(\phi_1,\chi,\phi_1;q-p,p)
\right.\nonumber\\
&&+Q_\chi s_W^2\left\{
\frac{1}{2}-2 C_{24}-M_Z^2(C_{12}+C_{23})+m_\chi^2 C_0
\right\}(\chi,\phi_1,\chi;q-p,p) \nonumber
\\
&&\left.
-\frac{1}{2}(B_0+B_1)(\phi_1,\chi;p)
\right],
\end{eqnarray}
where $p$ and $q$ are momenta for muon in $\Delta g^\mu_{L,R }$(muon neutrino in $\Delta g^{\nu_\mu}_L$) 
and Z-boson, respectively.
Correction to $\mu \nu_\mu W$-vertex is written by
\begin{eqnarray}
\Delta\bar{\delta}_G&=& 
\frac{y_L^2}{8\pi^2}\left[ \sum_i |V_{2i}|^2
 C_{24}(\phi_1,s_i,\chi;-q,p)
\right.\nonumber
\nonumber \\
&& \left. 
- \frac{1}{4}\left\{(B_0+B_1)(\phi_1,\chi;p-q)+
  \sum_i |V_{2i}|^2 (B_0+B_1)(s_i,\chi;p)
 \right\} \right],
\end{eqnarray}
where $p$, $p-q$ and $q$ are momenta for muon, muon-neutrino, and
W-boson, respectively. Here we use approximation $q^2=0$.

\begin{figure}[ht]
\centerline{
\includegraphics[width=8.0cm,angle=0]{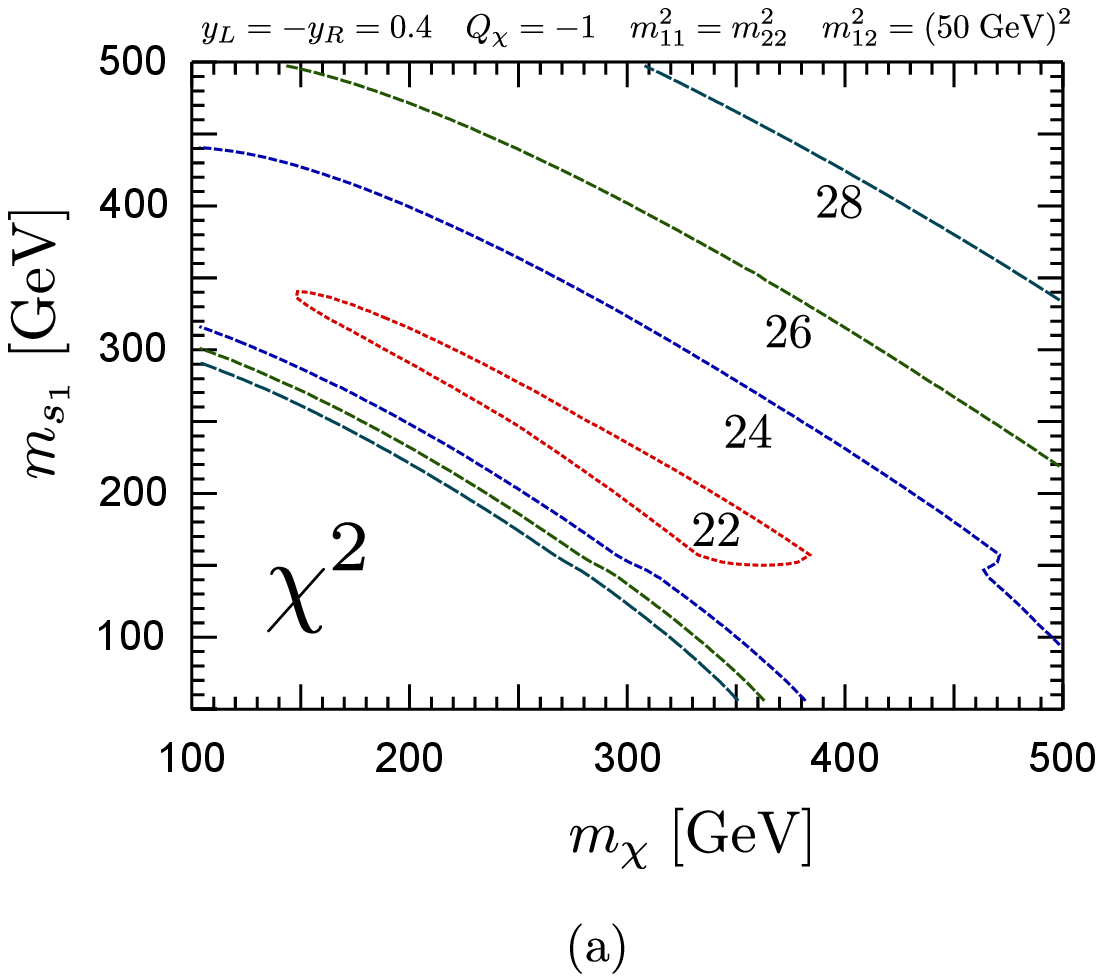}
\includegraphics[width=8.0cm,angle=0]{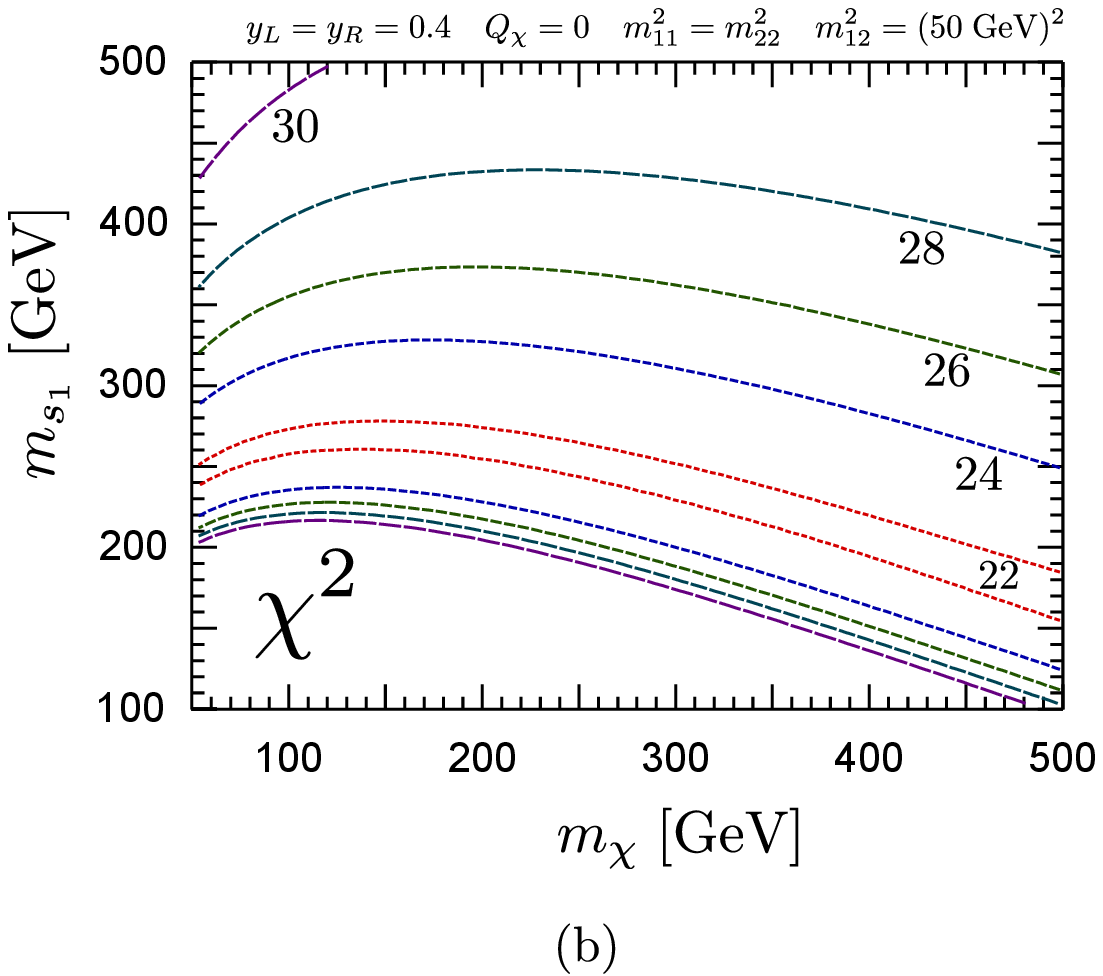}
}
\caption{$\chi^2$ contours as a function of $m_\chi$ and $m_{s_1}$ in case of
 (a) $Q_\chi=-1$ and $y_L=-y_R=0.4$ and (b) $Q_\chi=0$ and $y_L=y_R=0.4$. Here we take $m_{11}^2=m_{22}^2$ and
 $m_{12}^2=(50~{\rm GeV})^2$.}
\label{chi2_model2}
\end{figure}
In Fig.~\ref{chi2_model2}, we show $\chi^2$ contours as a function of $m_\chi$
and $m_{s_1}$ in case of (a) $Q_\chi=-1$ and $y_L=-y_R=0.4$ and (b) $Q_\chi=0$ and $y_L=y_R=0.4$. 
Here we have taken $m_{12}^2=(50~{\rm GeV})^2$,
and $m_{11}^2=m_{22}^2$. $m_{\phi_1}$ is chosen so that the $\chi^2$
becomes the smallest, provided that $m_{\phi_1}>105~{\rm GeV}$ for
$Q_\chi=-1$ and $m_{\phi_1}>46~{\rm GeV}$ for $Q_\chi=0$ respectively.
In both cases shown in Fig.~\ref{chi2_model2}, the $\chi^2$ can be as small as 22, which is
smaller than those in the previous case where only right-handed muon has new Yukawa coupling.
As seen from Fig.~\ref{chi2_model2}, the relatively light $\chi$ and $s_i$ are favored by the
EW observables. In Table.~1, we show the predicted values of the EW observables
in case of $Q_\chi=0$, $m_{\phi_1}=300$ GeV, $m_{\chi}=200$ GeV, $m_{11}^2=m_{22}^2=(250)^2$ GeV$^2$,
and $m_{12}^2=(50)^2$ GeV$^2$. As shown in Table.~1, the fit is quite good.
Therefore, it will be important to know what kind of effects we expect from these particles
at the LHC.
In a next section, we will discuss the possible effects of these new particles at the LHC.

\section{Phenomenology at the LHC}

In the previous section, we show that in order to explain the anomaly
of muon g-2, it is strongly suggested that there should be
the EW scale new particles. It is also suggested that multi-charged particles
may be favored by the anomaly of muon g-2. Therefore, these particles may be reachable directly 
and/or indirectly at the LHC.
Here we discuss the effects of these new particles in the Higgs decay $h\rightarrow \gamma \gamma$
and their direct productions at the LHC.

\subsection{$h\rightarrow \gamma \gamma$}
The SU(2) singlet and doublet scalars can couple to the Higgs boson through the following
scalar interactions:
\begin{eqnarray}
\cal{L}&=&-\kappa_1 \phi^\dagger \phi (H^\dagger H)-\kappa_2(\Phi^\dagger \Phi)(H^\dagger H)
-\kappa_3 (H^\dagger \Phi)(\Phi^\dagger H)\nonumber \\
&&-\kappa_4 M\left\{(H^\dagger \Phi)\phi^\dagger~{\rm +h.c.}
\right\},
\label{Higgs_int}
\end{eqnarray}
where $H$ is a standard model Higgs boson doublet.
In case where only SU(2) singlet scalar $\phi$ exists, only coupling $\kappa_1$ is non-zero, but in case
where both SU(2) singlet $\phi$ and doublet $\Phi$ scalars are there, all couplings are allowed\footnote{
When the QED charge of $\phi$ ($Q_\phi$) is zero,  terms such as $(H^\dagger \Phi)\phi$ and $(H^\dagger \Phi)^2$
are also possible. In our analysis, we assume that such terms are small, for simplicity. Even if we include
such terms, our result does not change qualitatively.
}.

In the case where only right-handed muon has new Yukawa interaction, there is SU(2) singlet $\phi$, and its interaction
with Higgs boson is
\begin{eqnarray}
 {\cal{L}}=-\lambda v h \phi^\dagger \phi,
\end{eqnarray}
where $\lambda=\kappa_1$ in this model and $v$ is a vacuum expectation value of Higgs field, $v\simeq 246$ GeV.
When $Q_\phi$, which is the QED charge of $\phi$, is non-zero, $\phi$ can contribute to the Higgs decay 
$h\rightarrow \gamma \gamma$.
The decay width $\Gamma(h\rightarrow \gamma \gamma)$ is given by
\begin{eqnarray}
 \Gamma(h\rightarrow \gamma \gamma)&=&\frac{\alpha^2 m_h^3}{256\pi^3 v^2} |S(m_h)|^2,
\end{eqnarray}
where the amplitude $S(m_h)$ at leading order is written by
\begin{eqnarray}
 S(m_h)=\frac{8}{3}F_t(\tau_t)-F_W(\tau_W)+Q_\phi^2 \lambda\frac{v^2}{2m_\phi^2}F_\phi(\tau_\phi).
\end{eqnarray}
Here the first, the second, and the third terms are top, W, and $\phi$ contributions, respectively, and
$\tau_x=\frac{m_h^2}{4m_x^2}~(x=t,W,~\phi)$. The function $F_t,~F_W,~F_\phi$ are given by
\begin{eqnarray}
 F_t(\tau)&=&\tau^{-1}\left[
1+(1-\tau^{-1})f(\tau)
\right],\\
F_W(\tau)&=&2+3\tau^{-1}+3\tau^{-1}(2-\tau^{-1})f(\tau),\\
F_\phi(\tau_\phi) &=& \tau^{-1}\left[-1+\tau^{-1}f(\tau)\right],
\end{eqnarray}
where $f(\tau)=\arcsin^2(\sqrt{\tau})$ for $\tau<1$. 
Branching ratio of $h\rightarrow \gamma \gamma$ is approximately given by
\begin{eqnarray}
 \frac{{\rm BR}(h\rightarrow \gamma \gamma)}{{\rm BR}(h\rightarrow \gamma \gamma)_{\rm SM}}
&\simeq&
 \frac{\Gamma(h\rightarrow \gamma \gamma)}{\Gamma(h\rightarrow \gamma \gamma)_{\rm SM}}
=\left|
\frac{\frac{8}{3}F_t(\tau_t)-F_W(\tau_W)+Q_\phi^2\lambda \frac{v^2}{2m_\phi^2}F_\phi(\tau_\phi)}
{\frac{8}{3}F_t(\tau_t)-F_W(\tau_W)}
\right|^2,
\label{Hgg}
\end{eqnarray}
because the total decay width does not change much, even if $\Gamma(h\rightarrow \gamma \gamma)$ changes.
Here ${\rm BR}(h\rightarrow \gamma \gamma)_{\rm SM}$ and $\Gamma(h\rightarrow \gamma \gamma)_{\rm SM}$ are
the SM predictions at the leading order.
\begin{figure}[ht]
\centerline{
\includegraphics[width=10.0cm,angle=0]{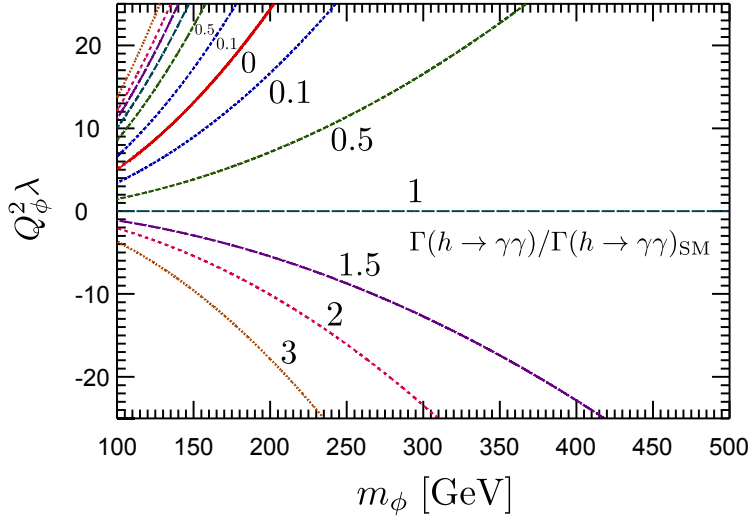}
}
\caption{Higgs decay width $h\rightarrow \gamma \gamma$ normalized by the SM prediction
$\frac{\Gamma(h\rightarrow \gamma \gamma)}{\Gamma(h\rightarrow \gamma \gamma)_{\rm SM}}$
as a function of $m_\phi$ and $Q_\phi^2\lambda$. Here we take $m_h=125$ GeV.
}
\label{Higgs_LHC}
\end{figure}
Notice that this ratio is a function of $m_\phi$ and $Q_\phi^2 \lambda$ when Higgs boson 
mass $m_h$ is fixed. As we discussed earlier, in order to explain the anomaly of
muon g-2, larger $Q_\phi$ may be favored. As can be seen in Eq.~(\ref{Hgg}), larger $Q_\phi$
induces larger effect on $h\rightarrow \gamma \gamma$.

In Fig.~\ref{Higgs_LHC}, the ratio 
${\Gamma(h\rightarrow \gamma \gamma)}/{\Gamma(h\rightarrow \gamma \gamma)_{\rm SM}}$
is shown as a function of $m_\phi$ and $Q_\phi^2\lambda$. Here we took $m_h=125$ GeV.
As one can see from the figure,
the effect can be significant if $Q_\phi^2\lambda$ is large and $m_\phi$ is $O(100)$ GeV.
For example, as we discussed, when $Q_\phi=4$, $\lambda=1~(-1)$ and $m_\phi=200$ GeV, 
${\Gamma(h\rightarrow \gamma \gamma)}/{\Gamma(h\rightarrow \gamma \gamma)_{\rm SM}}=0.12~(2.7)$.
Unfortunately, we can not predict the branching ratio ${\rm BR}(h\rightarrow \gamma \gamma)$
in this model because $\lambda$ is unknown. However, it is not surprising that 
${\rm BR}(h\rightarrow \gamma \gamma)$ is very different from the SM prediction.

In the case where both right- and left-handed muons have new Yukawa interactions, the SU(2) singlet
and doublet scalars have interactions with Higgs boson, as shown in Eq.~(\ref{Higgs_int}) and they 
generates the following interactions:
\begin{eqnarray}
 {\cal L} &=&-\kappa_2 v h\phi_1^*\phi_1-\sum_{ij} \lambda_{ij} vh s_i^* s_j, 
\end{eqnarray}
where
\begin{eqnarray}
 \lambda_{ij}=\kappa_1 V_{1i}^* V_{1j}+(\kappa_2+\kappa_3)V_{2i}^* V_{2j}
+\frac{M}{\sqrt{2}v}\kappa_4 (V_{1i}^* V_{2j}+V_{2i}^* V_{1j}).
\end{eqnarray}
They contribute to the decay width of $h\rightarrow \gamma \gamma$,

\begin{eqnarray}
&& \frac{{\rm BR}(h\rightarrow \gamma \gamma)}{{\rm BR}(h\rightarrow \gamma \gamma)_{\rm SM}}
\nonumber \\
\simeq&&
\left|
\frac{\frac{8}{3}F_t(\tau_t)-F_W(\tau_W)
+Q_1^2\kappa_2 
\frac{v^2}{2m_{\phi_1}^2}F_\phi(\tau_{\phi_1})
+Q_2^2
\sum_i
\lambda_{ii}
\frac{v^2}{2m_{s_i}^2}F_\phi(\tau_{s_i})
}
{\frac{8}{3}F_t(\tau_t)-F_W(\tau_W)}
\right|^2.
\label{Hgg_v2}
\end{eqnarray}

When $Q_\chi=-1$, which corresponds to $Q_1=1$ and $Q_2=0$
(as an example we discussed above), only $\phi_1$ can contribute
to $h\rightarrow \gamma \gamma$, because $s_i$ does not couple to photon.
The result is same as one in Fig.~\ref{Higgs_LHC} for $\lambda=\kappa_2$ and $Q_\phi=Q_{1}=1$.
In this case, the effect can be significant if $\kappa_2$ is large. For example,
when $m_{\phi_1}=250$ GeV and $\kappa_2=1$ and $-1$, 
${\Gamma(h\rightarrow \gamma \gamma)}/{\Gamma(h\rightarrow \gamma \gamma)_{\rm SM}}=0.95$ 
and 1.05, respectively.

When $Q_\chi=0$, which corresponds to $Q_1=0$ and $Q_2=-1$, scalars $s_i~(i=1,2)$ can
contribute to $h\rightarrow \gamma \gamma$. Similarly to the previous case,
if the Higgs couplings $\lambda_{11}$ and $\lambda_{22}$ are large, the effect
on $h\rightarrow \gamma \gamma$ can be significant.
For example, when $m_{11}^2=m_{22}^2$ and $M=\frac{v}{\sqrt{2}}$, we obtain $\lambda_{11}=\kappa$ and
$\lambda_{22}=2\kappa$ for $\kappa_i=\kappa >0~(i=1-4)$,
on the other hand, $\lambda_{11}=2\kappa$ and
$\lambda_{22}=\kappa$ for $\kappa_i=\kappa <0~(i=1-4)$.
As a result, for $m_{11}^2=m_{22}^2=(300~{\rm GeV})^2$,
${\Gamma(h\rightarrow \gamma \gamma)}/{\Gamma(h\rightarrow \gamma \gamma)_{\rm SM}}=0.90$ and
$1.14$ for $\kappa=1$ and $-1$ (0.71 and 1.56  for $\kappa=2$ and $-2$
), respectively. 
Although it is difficult to predict the decay branching ratio of
$h\rightarrow \gamma \gamma$ in this model because it depends on many unknown parameters
in scalar interactions, it is possible to have large effect on $h\rightarrow \gamma \gamma$
process.

\subsection{Direct productions at the LHC}

As we have discussed in the previous sections, the anomaly of muon
g-2 strongly suggests the existence of EW scale new particles.
Models we discussed in this paper may not be complete, namely, there may be
other new particles in the complete models. Therefore, in this paper, we can not discuss
the well-defined signatures at the LHC. However, we think that it is important
to know how many events of new particles can be generated at the LHC.
Here we present the production cross sections of new particles at the LHC.

\begin{figure}[ht]
\centerline{
\includegraphics[width=8.0cm,angle=0]{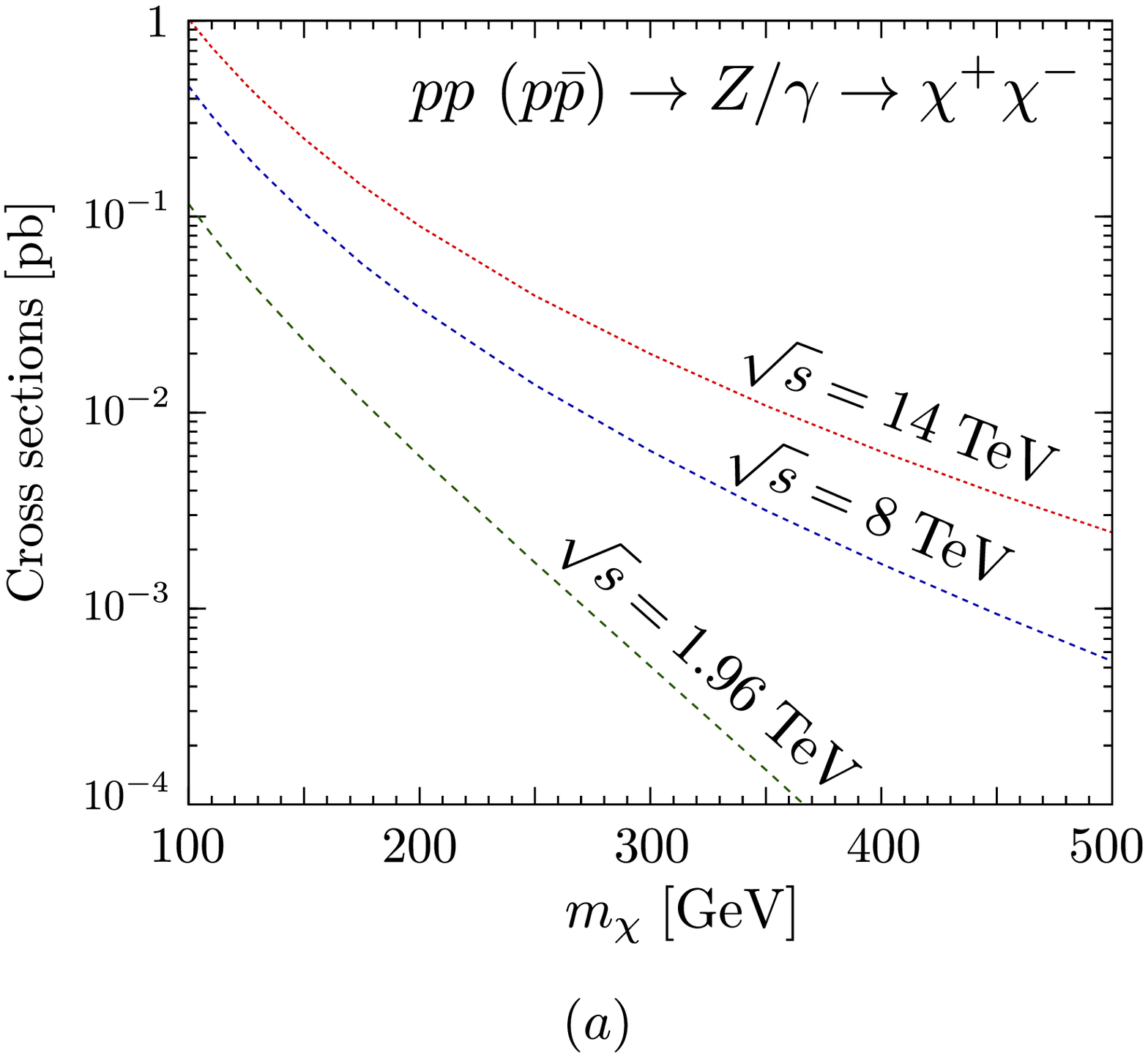}
\includegraphics[width=8.0cm,angle=0]{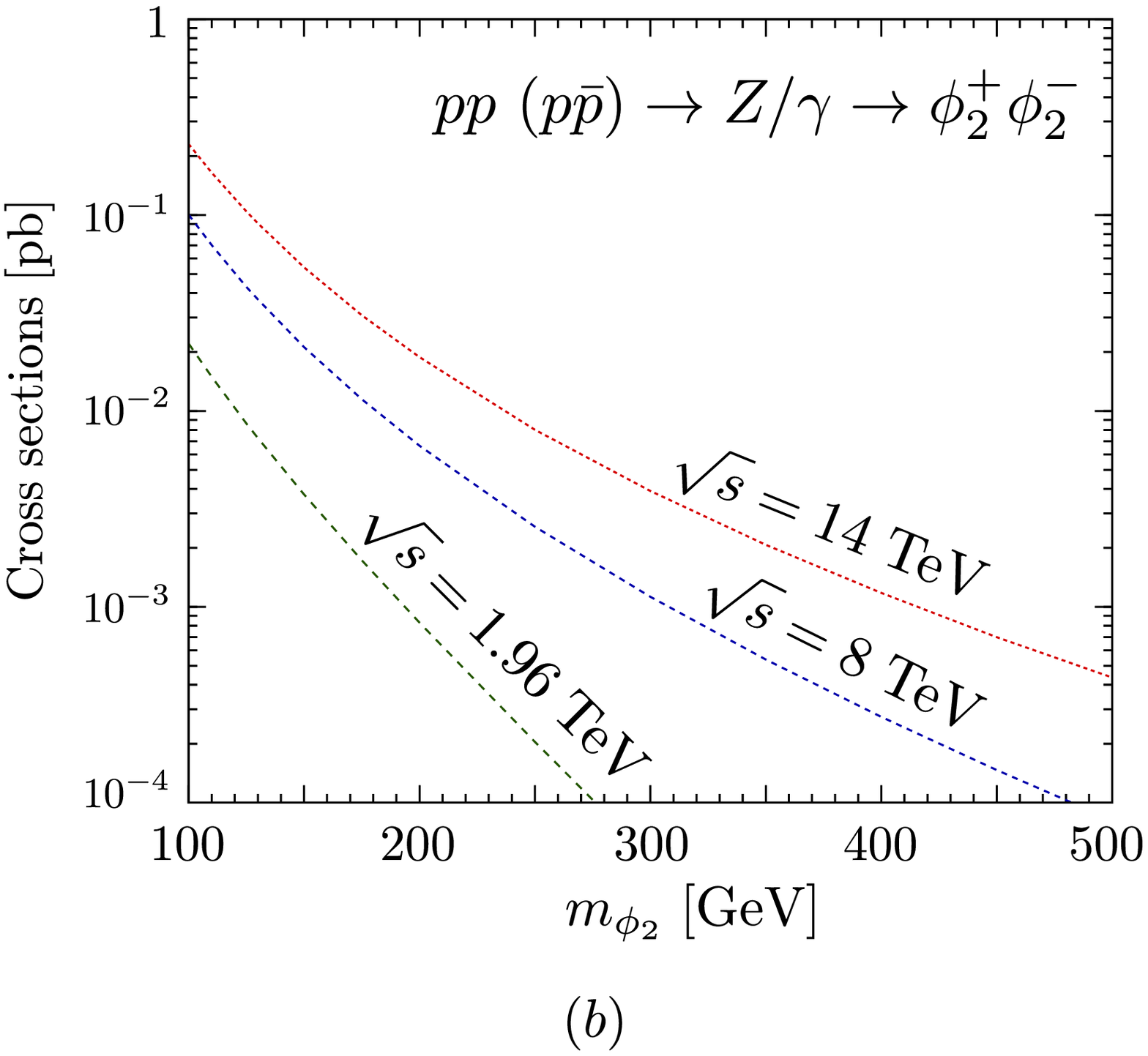}
}
\caption{Pair production cross sections of (a) $\chi$ in case of $Q_\chi=-1$:~
$pp~ (p\bar{p})\rightarrow Z/\gamma \rightarrow
 \chi^+\chi^-$, and 
(b) $\phi_2^-$ in case of $Q_2=-1$:~ $pp~ (p\bar{p})\rightarrow Z/\gamma \rightarrow
 \phi_2^+\phi_2^-$ at the LHC (Tevatron), as a function of $m_\chi$ and $m_{\phi_2}$, respectively.
Here we take $\sqrt{s}=8$ TeV and 14 TeV for LHC and 1.96 TeV
for Tevatron.
}
\label{fig_LHC}
\end{figure}

In the case where the right-handed muon has new Yukawa coupling, 
singlet scalar ($\phi$) and singlet fermion ($\chi$) exist in the model.
When $Q_\chi=-1$,  only $\chi$ couples to the standard model gauge bosons
($Z$ and $\gamma$) because $Q_\phi=0$. Therefore, the fermion $\chi$ can be pair-produced
via EW gauge interactions. 
We calculate the pair production cross section of $\chi$ at the LHC
($\sqrt{s}=8~{\rm TeV}~{\rm and}~14~{\rm TeV}$) and at Tevatron ($\sqrt{s}=1.96$ TeV)
as a function of $m_\chi$, as shown in Fig.~\ref{fig_LHC}(a). To
calculate the production cross sections, we have used the MadGraph~\cite{Alwall:2007st}.
In order to explain the anomaly of muon g-2, the mass of the fermion
$\chi$ should be of order of about 100 GeV. From Fig.~\ref{fig_LHC}(a), the
production cross section of $\chi$ is in the range of
0.5-0.0005 pb (1-0.001 pb) for $m_\chi=100-500$ GeV and $\sqrt{s}=8$ TeV ($\sqrt{s}=14$ TeV) 
at the LHC and also 0.1-0.001 pb for $m_\chi=100-250$ GeV and $\sqrt{s}=1.96$ TeV at the Tevatron.
If we consider a luminosity of 10 fb$^{-1}$ for $\sqrt{s}=8$ TeV and 14 TeV, the number of events produced 
at the LHC is of order of 5-5000 and 10-10000, respectively. Therefore, number of signal events
can be significant. 
In this case, if $\phi$ is lighter than $\chi$, $\chi$ can decay to $\mu$ and $\phi$. If $\phi$ is
a stable neutral particle, the final signature may be a $\mu^+\mu^-$+ missing energy. This will suffer
from the standard model background such as $W^+W^-(\rightarrow \mu^+\mu^-)$ production whose
production rate is about 0.29 (0.51) pb for $\sqrt{s}=8$ TeV (14 TeV) at the LHC. Therefore, the detail study will
be very important to discuss the discovery potential of such a signal event.
However, the detail signature
will depend on the particle spectrum in the complete model. Thus we will leave the detail study.
If we consider multi-charged $\chi$, the cross section increases by a factor 
$Q_\chi^2$, and the cross section further becomes larger, and hence the constraints from the LHC
will be very important.
\footnote{In addition, $\phi$ pair production is also 
not negligible in this case.}

In the second model where one has a $SU(2)_L$ doublet scalar
in addition to $\phi$ and $\chi$, the scalars can
be pair-produced via EW gauge interaction. For example, Fig.~\ref{fig_LHC} (b) shows
the production cross sections for the charged
$\phi_2^+ \phi_2^-$ scalars as a function of $m_{\phi_2}$
in case of $Q_\chi=0$, which corresponds to $Q_1=0$ and $Q_2=-1$. We show the predicted values
for $\sqrt{s}=8$ TeV and 14 TeV at the LHC and $\sqrt{s}=1.96$ TeV at the Tevatron.
As can be seen from the figure, the cross sections 
are in the range of
0.1-0.0001 pb (0.2-0.0005 pb) for $m_{\phi_2}=100-500~$ GeV at $\sqrt{s}=8$ TeV ($\sqrt{s}=14$ TeV)
and also  0.02-0.0001 pb for for $m_{\phi_2}=100-250$ GeV at the Tevatron.
For simplicity, we have assumed that there are no $\phi$-$\phi_2$ scalar mixing.
In this case, $\phi_1+\phi_2$ are also produced via $W$-boson.
If masses of $SU(2)_L$ doublet scalar are degenerated, we have checked that the cross section
is similar to one in Fig.~\ref{fig_LHC} (b). In general,
the signal events have the form $pp~ (p\bar{p})\to V\to \phi^*_i\phi_j$ where
$V=Z,\gamma,W$ and $i,j=1,2$. Therefore, Fig.~\ref{fig_LHC} shows the typical values 
of the cross section in this model.

Although we can not discuss the complete signatures because our sample models are not
complete, we stress that the searches for the production processes of these new particles via
EW interactions are very important in models where the anomaly of muon g-2
can be explained by these new particles.

\section{Summary}
The current LHC experiment does not show any deviation from the SM predictions.
This result puts severe constraints on new physics models. Especially, colored particles
are strongly constrained since the LHC as a hadron collider can easily produce the colored
particles via QCD interactions. For example, in the constrained minimal supersymmetric standard
model (CMSSM), gluino as well as the first and the second generation squarks are strongly
constrained and their typical lower mass limits are about 1 TeV~\cite{Aad:2012hm}. Therefore, models
motivated by the hierarchy problem are getting more constrained, and especially 
CMSSM as a solution to
the hierarchy problem is strongly disfavored.

If we motivate new physics models from the reported anomaly of muon g-2, our naive
expectation is the existence of EW scale new particles because the size of anomaly of
muon g-2 is as large as one induced by EW gauge bosons in the SM.
In this paper, we analyzed two examples, where muon has new Yukawa type interactions
with new particles. One example was a model where right-handed muon has new Yukawa interaction
with SU(2) singlet scalar and fermion. In this case, in order to explain the anomaly of 
muon g-2 as well as to satisfy the precision EW measurements, we found that the relatively 
large new Yukawa coupling and EW scale new particles are strongly favored.
We also noted that the multi-charged particles (whose QED charges are large) are favored.

Another example we analyzed was a model where both right- and left-handed muons have
new Yukawa interactions with SU(2) doublet scalar and singlet scalar and fermion, so that the chirality
flip of muon can occur in the internal fermion line of 1-loop muon g-2 diagrams.
In this case, new contributions to muon g-2 are enhanced, and hence even smaller new Yukawa
couplings can explain the anomaly of muon g-2. We also showed that the EW observables 
are also well-fitted by the new physics contributions. In this case, the EW scale new particles
are favored by the data.

Although these two examples may be only a part of complete models, the muon g-2
as well as the EW observables can constrain the QED charge of new particles as well 
as the scale  of new particles and the size of new Yukawa couplings.
Since the EW scale new particles can explain the anomaly of
muon g-2 and they can be consistent with the precision EW measurements, it is important
to study the possible phenomenology at the LHC.

Since some of new particles have QED charges as well as couplings with Higgs boson, the
branching ratio of $h\rightarrow \gamma\gamma$ can be effected. Especially, as we showed,
multi-charged particles favored by the muon g-2 can significantly
contribute to $h\rightarrow \gamma\gamma$, and the effects can be sizable. Therefore,
Higgs physics will have an impact on models we discussed here.

We also computed direct production cross sections of new particles motivated by
the muon g-2. Since models discussed here may not be complete, we could not
discuss the complete signatures for the models. However, since the number of signal events
are not negligible, there may be possibility to reveal new physics models for muon g-2 at the LHC.
Since we do not need the colored particles to explain the anomaly of muon g-2, the search for
new particles which only have EW interactions will be very important.

Finally, we comment on the possible completions of models discussed in this paper.
The structure of the model where both right- and left-handed muons have new Yukawa 
type interactions is quite similar to that of the minimal supersymmetric standard model.
When $Q_2=-1$, the scalar $\phi_2$ in SU(2) doublet scalar $\Phi$ and
SU(2) singlet scalar $\phi$ correspond to left- and right-handed smuons, respectively.
The SU(2) singlet fermion $\chi$ is a Bino, for example. 
A part of a model discussed in Ref.~\cite{Hambye:2006zn}, which is motivated by neutrino mass 
and dark matter is also similar to the model we discussed here. Therefore, our approach
can capture crucial features of new physics models for the anomaly of muon g-2 and hopefully 
create an interesting way
to find the physics beyond the standard model.

{\it Note added}

After we completed this work, we learned that ATLAS and CMS collaboration at LHC observed a
new particle consistent with the SM Higgs boson at a mass of about 125 GeV~\cite{:2012gk}.
Since we have taken the Higgs boson mass to be 125 GeV in our analysis,
our conclusion does not change.

\setcounter{equation}{0}
\setcounter{footnote}{1}

\addcontentsline{toc}{section}{Appendix: Useful functions}

\section*{Acknowledgments}
This work is supported in part by Grant-in-Aid for Scientific research from the
Ministry of Education, Science, Sports, and Culture (MEXT), Japan, 
No. 22540273 (KT). This work is also supported in part by Grant-in-Aid for 
Nagoya University Global COE Program, ``Quest for Fundamental Principles in the
Universe: from Particles to the Solar System and the Cosmos'', from the MEXT, Japan.

\newpage
\appendix
\section{Useful function}
In this paper, Passarino-Veltman functions~\cite{'tHooft:1978xw} are defined by
\begin{eqnarray}
A(A)&=&
16\pi^2\mu^{2\epsilon}\int
\frac{d^n k}{i(2\pi)^n}\frac{1}{k^2-m_A^2+i\epsilon},
\\
B_0(A,B;p) &=&
16\pi^2\mu^{2\epsilon}\int
\frac{d^nk}{i(2\pi)^n}\frac{1}{\left[k^2-m_A^2+i\epsilon\right]
\left[(k+p)^2-m_B^2+i \epsilon\right]},\nonumber
\\
p^\mu B_1(A,B;p) &=& 16\pi^2 \mu^{2\epsilon}
\int \frac{d^n k}{i(2\pi)^n}\frac{k^\mu}
{\left[k^2-m_A^2+i\epsilon\right]\left[(k+p)^2-m_B^2+i\epsilon\right]},
\nonumber
\\
p^\mu p^\nu B_{21}(A,B;p)&+&g^{\mu \nu}B_{22}(A,B;p)
\nonumber \\
&=& 16\pi^2 \mu^{2\epsilon} \int
\frac{d^n k}{i(2\pi)^n} \frac{k^\mu k^\nu}
{\left[k^2-m_A^2+i\epsilon\right]\left[(k+p)^2-m_B^2+i\epsilon\right]},
\end{eqnarray}
\begin{eqnarray}
&& C_0(A,B,C;p_1,p_2)\nonumber \\
&=&
16\pi^2 \mu^{2\epsilon}
\int \frac{d^n k}{i(2\pi)^n}\frac{1}
{[k^2-m_A^2+i\epsilon][(k+p_1)^2-m_B^2+i\epsilon][(k+p_1+p_2)^2-m_C^2+i\epsilon]},
\nonumber 
\\
&&\left(
p_1^\mu C_{11}+p_2^\mu C_{12}
\right)(A,B,C;p_1,p_2) \nonumber \\
&=& 16\pi^2 \mu^{2\epsilon}
\int \frac{d^n k}{i(2\pi)^n}\frac{k^\mu}
{[k^2-m_A^2+i\epsilon][(k+p_1)^2-m_B^2+i\epsilon][(k+p_1+p_2)^2-m_C^2+i\epsilon]},
\nonumber 
\\
&&\left\{(p_1^\mu p_1^\nu C_{21}
+p_2^\mu p_2^\nu C_{22}+(p_1^\mu p_2^\nu+p_1^\nu p_2^\mu)C_{23} +g^{\mu \nu}C_{24}
\right\}
(A,B,C;p_1,p_2) \nonumber \\
&=& 16\pi^2 \mu^{2\epsilon}
\int \frac{d^n k}{i(2\pi)^n}\frac{k^\mu k^\nu}
{[k^2-m_A^2+i\epsilon][(k+p_1)^2-m_B^2+i\epsilon][(k+p_1+p_2)^2-m_C^2+i\epsilon]},
\nonumber 
\\
\end{eqnarray}
where we use dimensional regularization in $4-2\epsilon$ dimensions.

For convenience, we list explicit forms of some of above functions.
\begin{eqnarray}
A(m^2)&=&
m^2\left(
\frac{1}{\Delta}+1-\log\frac{m^2}{\mu^2}
\right),\\
B_0(A,B;p)&=& \frac{1}{\Delta}-\int_0^1 dx \log
\frac{m_1^2(1-x)+m_2^2x -p^2x(1-x)-i\epsilon}{\mu^2},\\
B_1(A,B;p)&=&
-\frac{1}{2\Delta}+\int_0^1 dx x\log
\frac{m_1^2(1-x)+m_2^2x -p^2x(1-x)-i\epsilon}{\mu^2},\\
B_{21}(A,B;p)&=&
\frac{1}{3\Delta}-\int_0^1 dx x^2\log
\frac{m_1^2(1-x)+m_2^2x -p^2x(1-x)-i\epsilon}{\mu^2},\\
B_{22}(A,B;p)&=&
\frac{1}{4}(m_1^2+m_2^2-\frac{p^2}{3})\left(\frac{1}{\Delta}+1\right)
\nonumber \\
&&\hspace{-4cm}
-\frac{1}{2}\int_0^1dx \left\{
m_1^2(1-x)+m_2^2 x-p^2x(1-x)
\right\}\log
\frac{m_1^2(1-x)+m_2^2x -p^2x(1-x)-i\epsilon}{\mu^2},
\end{eqnarray}
where $\frac{1}{\Delta}=\frac{1}{\epsilon}-\gamma+\log 4\pi$.

\end{document}